\documentclass[1p]{elsarticle}   

\usepackage{lineno,hyperref}
\usepackage{multirow}
\usepackage{xcolor}
\usepackage{amsmath}

\modulolinenumbers[2]

\journal{J. Quant. Spectrosc. Radiat. Transfer}

\biboptions{sort&compress}

\begin{document}

\begin{frontmatter}

\title{Spectroscopic line parameters of NO, NO$_2$, and N$_{2}$O for the HITEMP database}

\author[firstaddress]{Robert J. Hargreaves\corref{corresponding_author}}
\cortext[corresponding_author]{Corresponding author} \ead{robert.hargreaves@cfa.harvard.edu}

\author[firstaddress]{Iouli E. Gordon}
\author[firstaddress]{Laurence S. Rothman}
\address[firstaddress]{Center for Astrophysics \textbar Harvard \& Smithsonian,  Atomic and Molecular Physics Division, 60 Garden Street, Cambridge, MA 02138, USA}

\author[secondaddress]{Sergey A. Tashkun}
\author[secondaddress]{Valery I. Perevalov}
\author[secondaddress]{Anastasiya A. Lukashevskaya}
\address[secondaddress]{V. E. Zuev Institute of Atmospheric Optics, Siberian Branch, Russian Academy of Sciences, Tomsk 634055, Russia}

\author[thirdaddress]{Sergey N. Yurchenko}
\author[thirdaddress]{Jonathan Tennyson}
\address[thirdaddress]{Department of Physics and Astronomy, University College London, London, WC1E 6BT, UK}

\author[fouthaddress]{Holger S. P. M\"{u}ller}
\address[fouthaddress]{I. Physikalisches Institut, Universit\"{a}t zu K\"{o}ln, Z\"{u}lpicher Str. 77, 50937 K\"{o}ln , Germany}

\begin{abstract}
This work describes the update of NO along with the incorporation of NO$_{2}$ and N$_{2}$O to the HITEMP database. Where appropriate, the HITRAN line lists for the same molecules have also been updated. This work brings the current number of molecules provided by HITEMP to seven. The initial line lists originating from \textit{ab initio} and semi-empirical methods for each molecule have been carefully validated against available observations and, where necessary, adjustments have been made to match observations. We anticipate this work will be applied to a variety of high-temperature environments including astronomical applications, combustion monitoring, and non-local thermodynamic equilibrium conditions.

\end{abstract}

\begin{keyword}
High-temperature spectroscopy\sep NO \sep NO$_{2}$ \sep N$_{2}$O \sep line lists \sep HITEMP

\end{keyword}

\end{frontmatter}




\section{Introduction}\label{introduction}

Nitrogen oxides play a key role in atmospheric chemistry. Nitric oxide (NO) and nitrogen dioxide (NO$_{2}$), collectively referred to as NO$_{x}$, are a primary source of air pollution and contribute to the formation of smog and acid rain \cite{1983JGR....8810785L}. Nitrous oxide (N$_{2}$O), while not considered a strict member of the NO$_{x}$ family, is a major greenhouse gas and has recently overtaken chlorofluorocarbons (CFCs) as the dominant anthropogenic cause of ozone depletion \cite{2009Sci...326..123R}. Additional nitrogen oxides can be observed in the terrestrial atmosphere; however their abundance is significantly smaller and not considered here.

Atmospheric NO$_{x}$ is primarily a consequence of the combustion of fossil fuels at high temperatures. However, NO$_{x}$ is also readily produced by other means, including the burning of biomass material (natural and man-made), soil emissions \cite{2008JGRD..113.4302V}, and lightning strikes \cite{1997JGR...102.5929P}.

NO$_{x}$ gases are highly poisonous and detrimental to air quality. For this reason, NO$_{x}$ gases are heavily monitored and regulated by governments (e.g., via the Clean Air Act in the USA) and can be measured in human breath samples when monitoring health effects \cite{Heinrich2009}. Diesel engines are notorious for the levels of NO$_{x}$ they produce during the combustion process, since the formation of NO$_{x}$ is strongly coupled to combustion temperatures \cite{2005..book..M}. The monitoring of NO$_{x}$ emissions is therefore a crucial aspect of the automotive industry. A recent international scandal centered on the under-reporting of diesel engine emissions: the highlighted NO$_{x}$ emissions exceeded legal levels by up to 40 times \cite{Barrett_2015}. The spectroscopy of NO$_{x}$ has been applied to non-invasive thermometry of high-enthalpy environments \cite{2014MeScT..25l5103S} and \textit{in situ} measurements of the exhaust gases of coal-fired power plants \cite{CHAO2011725} and vehicles \cite{2011ApPhB.105..923Y}.

In addition to combustion environments, NO is also present in observations of nightglow in the atmospheres of Earth \cite{1975GeoRL...2..179G}, Mars \cite{2005Sci...307..566B}, and Venus \cite{2009PNAS..106..985G}. Terrestrial emission from the thermosphere has also shown NO to be in non-local thermodynamic equilibrium (NLTE) \cite{2007JGRA..11210301G}, requiring the consideration of high rotational transitions in limb-view radiative transfer models. N$_{2}$O, commonly referred to as laughing gas (and used worldwide as an anaesthetic), can be employed as a component for rocket propulsion \cite{2001AcAau..48..353Z}, remaining relatively stable up to 1100~K.

Additionally, NO$_{x}$ gases have been predicted to be present in the atmospheres of Earth-like exoplanets \cite{2017MNRAS.470..187A}, with  N$_{2}$O proposed as a potential biosignature \cite{2018AsBio..18..663S}.

\subsection{The HITRAN and HITEMP databases}\label{intro_hitran}

The HITRAN database contains detailed molecular spectroscopic parameters of 49 molecules (along with absorption cross-sections, collision-induced absorption spectra and data of aerosol properties), with HITRAN2016  \cite{2017JQSRT.203....3G} the most recent version. It is freely available at HITRAN\textit{online}\footnote{\href{https://hitran.org}{https://hitran.org}}, and includes spectroscopic parameters of NO, NO$_{2}$ and N$_{2}$O. The HITRAN database is used as an input source for radiative-transfer codes in a variety of applications, but the principal use for HITRAN is in the remote sensing of the terrestrial atmosphere. The parameters have a reference temperature of 296~K, and a description of all HITRAN parameters is given in Ref. \cite{2017JQSRT.203....3G}. The HITRAN data are extremely successful at modeling spectra in the temperature regime of the terrestrial atmosphere. Recent updates have expanded HITRAN to include additional broadening parameters for H$_{2}$, He, and CO$_{2}$ of relevance to the atmospheres of Jupiter, Saturn and Mars \cite{2016JQSRT.168..193W}. However, application of HITRAN to environments with elevated temperatures can lead to an incomplete model. This incompleteness is a consequence of HITRAN containing predominantly rotational, vibrational and electronic transitions that are relevant for the terrestrial atmosphere (and typically observed). For this reason, it is not necessary to include some high rotational transitions, many vibrational hot bands, or extremely weak transitions since they have no observable effect to the spectrum (at terrestrial atmospheric temperatures). However, as the temperature increases, the distribution of populated energy levels changes such that these `weak' transitions can no longer be ignored.

The original HITEMP database \cite{1995SPIE.2471..105R} was established to be a molecular spectroscopic database specifically aimed at modeling gas phase spectra for high-temperature applications. Improvements in the accuracy of theoretical calculations for high temperature  (e.g., Ref. \cite{2006MNRAS.368.1087B}) enabled significant advancements for the characterization of high-temperature environments (e.g., exoplanet atmospheres \cite{2007Natur.448..169T}). These studies prompted a comprehensive update to the HITEMP database, leading to HITEMP2010  \cite{2010JQSRT.111.2139R}, for which the most reliable data was collated for five molecules: H$_{2}$O, CO$_{2}$, CO, NO and OH.

The most noticeable difference between the HITRAN and HITEMP databases is the number of lines included for each molecule. As an example, the principal isotopologue of H$_{2}$O currently has 141,360 lines when downloaded from HITRAN\textit{online}, as opposed to 114,241,164 in HITEMP2010: an $\sim$800-fold increase (both line lists cover the same wavenumber range of 0 to almost 30,000 cm$^{-1}$). This increase in the number of transitions will also increase the burden on line-by-line calculations by a similar order. These (numerous) additional lines are not necessary for terrestrial atmospheric applications and are not included in HITRAN.

Where possible, the HITRAN data are consistent with HITEMP. However, providing consistency between HITRAN and HITEMP is a challenge for molecules with numerous experimental and theoretical reference sources, and often involves substituting the HITRAN data into these (extremely large) theoretical line lists: a non-trivial task. Differing theoretical methods and mixing of modes can mean that quantum number assignments between sources are not necessarily comparable for high polyads, and unique line identification becomes a challenge. For both HITRAN and HITEMP, particular attention is given to the validation of line lists and care is taken when merging data, but occasionally this method can introduce small discontinuities in line positions or intensities as the data sources change.

The line parameters included for HITEMP are described in Ref. \cite{2010JQSRT.111.2139R}, and are also provided at a reference temperature of 296~K. While this may seem low for a high-temperature database, the consistency with HITRAN helps to avoid confusion (for example when a simulation code uses both databases simultaneously such as in a scenario where a hot source is viewed through a layered atmosphere). Furthermore, the line intensities of both databases ($S_{if}$) can be routinely scaled to an alternate temperature ($T$), using
\begin{equation}\label{eqn_temperature}
\begin{split}
  S_{if}(T) = & S_{if}(T_{ref})\frac{Q(T_{ref})}{Q(T)} \exp\left( \frac{c_{2}E_{i}}{T_{ref}} - \frac{c_{2}E_{i}}{T} \right)  \left[ \frac{1 - \exp(-c_{2}\nu_{if}/T)}{1 - \exp(-c_{2}\nu_{if}/T_{ref})} \right] \; ,
\end{split}
\end{equation}

\noindent
where $T_{ref}=296$~K, $c_{2}$ is the second radiation constant (i.e., $c_{2} = hc/k = 1.43877$~cmK), $Q$ is the total partition sum, $E_{i}$ is the lower state energy and $\nu_{if}$ is the transition frequency (both in cm$^{-1}$). All quantities used in Eq.~\ref{eqn_temperature} are provided via the HITRAN and HITEMP databases.

Since the release of HITEMP2010, there have been a number of experimental, theoretical, and semi-empirical studies aimed at improving the high-temperature modeling of small molecules. This work describes the update of NO and and additions of NO$_{2}$ and N$_{2}$O to the HITEMP database. The molecules are described in order of their ID in HITRAN.




\section{Nitrous Oxide (molecule 4)}\label{mol_n2o}

\subsection{Description of N$_{2}$O line lists}\label{line_lists_n2o}
N$_{2}$O was not included as part of HITEMP2010 \cite{2010JQSRT.111.2139R}, and for the HITEMP update described in this work, only $^{14}$N$_{2}$$^{16}$O will be considered. This updated N$_{2}$O line list has been completed using current line parameters from HITRAN2016 \citep{2017JQSRT.203....3G} and the recent \textit{NOSD-1000} \citep{2016JQSRT.177...43T} line list. These line lists are briefly described. 

\subsubsection{N$_{2}$O in HITRAN2016}\label{list_hitran_n2o}
HITRAN2016 contains five isotopologues of N$_{2}$O: $^{14}$N$_{2}$$^{16}$O, $^{14}$N$^{15}$N$^{16}$O, $^{15}$N$^{14}$N$^{16}$O, $^{14}$N$_{2}$$^{18}$O, $^{14}$N$_{2}$$^{17}$O with relative abundances of 0.990333, 0.003641, 0.003641, 0.001986, 0.000369, respectively.  

The $^{14}$N$_{2}$$^{16}$O line list predominantly constitutes line positions and intensities from \citet{2004..web.Toth}. However, pure rotational transitions are included from \citet{1987ApOpt..26.4058R} and \citet{1994JQSRT..52..447C}. Infrared measurements of the $\nu_{2}$ band ($0110-0000$) near 17~$\mu$m (580~cm$^{-1}$) are from \citet{1996JMoSp.177..203J} and data for the $0220-0000$ and $0001-1000$ bands are taken from \citet{2002JQSRT..72...37D}. A small number of additional lines have been added that were absent from the \citet{2004..web.Toth} line list. In total, the HITRAN2016 $^{14}$N$_{2}$$^{16}$O line list constitutes 33,074 lines and covers the $0.838-7796.633$ cm$^{-1}$ range.

\subsubsection{The \textit{NOSD-1000}}\label{list_nosd1000_n2o}

The Nitrous Oxide Spectroscopic Databank at 1000 K (\textit{NOSD-1000}) includes only $^{14}$N$_{2}$$^{16}$O \cite{2016JQSRT.177...43T}. The purpose of \textit{NOSD-1000} is to have a line list capable of simulating the spectrum of $^{14}$N$_{2}$$^{16}$O at high-resolution, and at high temperatures. The production of the \textit{NOSD-1000} has been performed in a similar way to previous work on CO$_{2}$ \cite{2003JQSRT..82..165T} (which was incorporated into HITEMP2010 \citep{2010JQSRT.111.2139R}). \textit{NOSD-1000} has been calculated using an effective Hamiltonian and respective effective dipole moment operator. Their parameters have been fit to observed line positions and  intensities. The line widths have been calculated using a semi-empirical approach. The \textit{NOSD-1000} line list is available to download from the Institute of Atmospheric Optics\footnote{\href{ftp://ftp.iao.ru/pub/LTS/NOSD-1000}{ftp://ftp.iao.ru/pub/LTS/NOSD-1000}} and contains more than 1.4 million transitions covering the $260-8310$ cm$^{-1}$ spectral region, with an intensity cutoff at $1.0\times10^{-25}$ cm/molecule applied at 1000~K.

\subsection{Comparisons of \textit{NOSD-1000} and recalculation}\label{comp_nosd_n2o}

\begin{figure}[t!]
\centering
\includegraphics[scale=0.3, trim={0 0 0 0}, clip]{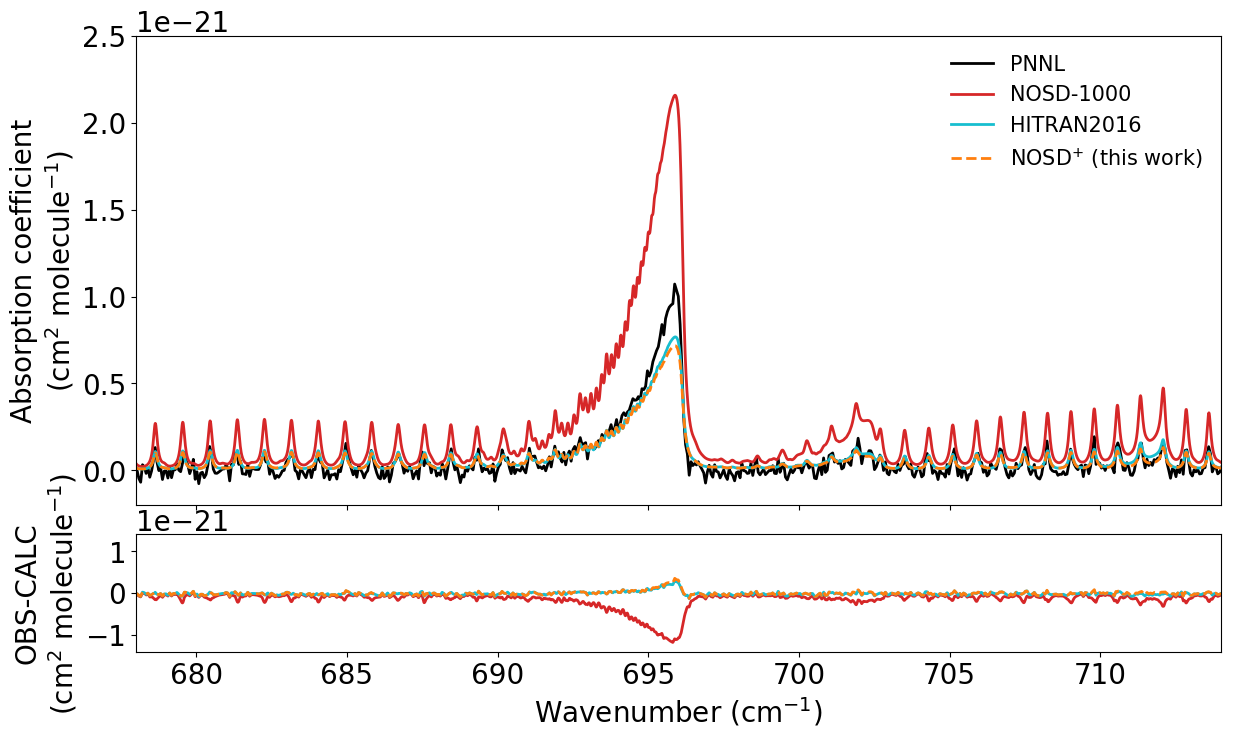}
\caption{Absorption cross sections of $^{14}$N$_{2}$$^{16}$O calculated using PNNL (black), \textit{NOSD-1000} (red), HITRAN2016 (blue),  and \textit{NOSD-1000$^{+}$} (orange, and included in this work) at 25$^{\circ}$C in the spectral region containing the 1000-0110 band. The top panel displays the absorption cross sections and the lower panel shows the residuals (i.e., PNNL-calc).  The spectra have been calculated using HAPI \cite{2016JQSRT.177...15K}. \label{n2o_1}}
\end{figure}

Broadband comparisons between absorption cross-sections calculated using \textit{NOSD-1000}, HITRAN2016, and Pacific Northwest National Laboratory (PNNL) \cite{2004ApSpe..58.1452S} line lists at 25$^{\circ}$C and $500 - 6500$ cm$^{-1}$ showed mostly excellent agreement. However, in the spectral region containing the $1000-0110$ (perpendicular $\nu_{1} \rightarrow \nu_{2}$) band, a large intensity discrepancy between \textit{NOSD-1000} and both HITRAN2016 and PNNL was observed (as shown in Fig.~\ref{n2o_1}). Furthermore, it was apparent that using the \textit{NOSD-1000} line list to calculate the N$_{2}$O spectra at lower temperatures would result in missing features, a consequence of the intensity cutoff applied at 1000~K.

The intensity error associated with the $1000-0110$ spectral region was improved for this work by correcting the effective dipole moment used to calculate intensities. This prompted a full re-calculation of the line list, to produce \textit{NOSD-1000b} along with \textit{NOSD-500b} and \textit{NOSD-296b} (calculated at 1000~K, 500~K and 296~K, respectively). The \textit{NOSD-1000b} line list contained a lower intensity cutoff, and was adjusted for each temperature. These corrected line lists are summarized in Table.~\ref{n2o_tab}. It should be noted that the lowering of the intensity cutoffs for these line lists improves the completeness at high temperatures. However, many interpolyad resonance interactions, which become more frequent for highly-excited states, have not been accounted for. As a result, these perturbed bands may display considerable shifts in line positions and, consequently, the line intensity between perturbed bands will be redistributed. These effects should not be observable for low resolution comparisons, but may be seen at high resolution.

\begin{table}[t!]
\begin{center}
\caption{Overview of \textit{NOSD-1000} and recalculated line lists after correction to the effective dipole moment.\label{n2o_tab} }
\begin{tabular}{lccr}
\hline
\vspace{-0.3cm}
 & & & \\
\multirow{ 2}{*}{Line list}  & Temperature & Intensity cutoff & Number  \\ \vspace{0.1cm}
        & (K) & (cm/molecule)     & of lines \\
\hline
\vspace{-0.3cm}
 & & & \\
\textit{NOSD-1000}  & 1000 & $1.0\times10^{-25}$ & 1,405,069 \\
\textit{NOSD-296b}  &  296 & $1.0\times10^{-27}$ &   164,153 \\
\textit{NOSD-500b}  &  500 & $1.0\times10^{-27}$ &   786,511 \\  \vspace{0.1cm}
\textit{NOSD-1000b} & 1000 & $1.0\times10^{-26}$ & 3,453,360 \\
\hline
\end{tabular}
\end{center}
\end{table}

\subsection{Combining line lists}\label{list_combined_n2o}

The three NOSD line lists for $^{14}$N$_{2}$$^{16}$O described in Sect.~\ref{comp_nosd_n2o} have been converted into HITRAN format and all intensities have been scaled to 296~K (using Eq.~\ref{eqn_temperature}). However, each individual list only remains applicable at the calculated temperature, since each intensity cutoff excluded strong lines at higher or lower temperatures. To provide an update for HITEMP, it was therefore necessary to combine these line lists into a merged version (\textit{NOSD$^{+}$}) that would be applicable over the full $296-1000$~K temperature range.

In order to merge these line lists, all 164,153 transitions at 296~K (i.e., contained in \textit{NOSD-296b}) are retained. The \textit{NOSD-500b} line list is then compared to \textit{NOSD-296b} and any lines which do not appear at 296~K (in this case, 622,774) are added to the merged list. Finally, \textit{NOSD-1000b} is compared to the merged list, and any line which does not appear in this list (2,712,285) are included. This \textit{NOSD$^{+}$} contains a total of 3,499,212 unique lines. It is worth noting that the \textit{NOSD-1000b} line list did not contain 45,852 lines that are strong at lower temperatures. These lines typically include low rotational levels of fundamental vibrational bands and equates to almost 28\% of the \textit{NOSD-296b} line list.

Due to the change in the intensity cutoff between the three line lists, the new \textit{NOSD$^{+}$} line lists remains complete for $^{14}$N$_{2}$$^{16}$O between $296-500$~K for an intensity $>1.0\times10^{-27}$ cm/molecule. Between 500~K and 1000~K the NOSD-1000$^{+}$ is complete for an intensity $>1.0\times10^{-26}$ cm/molecule, which will gradually scale toward $1.0\times10^{-27}$ cm/molecule at 500~K. An absorption cross section calculated using \textit{NOSD$^{+}$} at 25$^{\circ}$C is also included in Fig.~\ref{n2o_1}, showing the improvement over the previous \textit{NOSD-1000} list.

One aim of HITEMP is to remain consistent with HITRAN where appropriate. Since the data contained in HITRAN are considered to be the most accurate available, it was considered appropriate to retain these lines for the HITEMP update. Hence, lines that appear in HITRAN2016 replace the corresponding lines in \textit{NOSD$^{+}$}.

The infrared spectrum of N$_{2}$O can be divided into a polyad spectral structure since $\nu_{3} \approx 2200$ cm$^{-1}$  $\approx 2\nu_{1} \approx  4\nu_{2}$. For calculated line lists, it is well know that this causes a problem for assignments due to the heavy mixing of the three vibrational modes. Therefore, the HITRAN2016 lines have been matched to corresponding lines in \textit{NOSD$^{+}$} for positions to within $\pm0.01$ cm$^{-1}$, intensities to within a factor of 2, and lower-state energies also to within $\pm0.01$ cm$^{-1}$.

The HITRAN quantum number format for N$_{2}$O uses spectroscopic nomenclature ($v_{1}$, $v_{2}$, $l_{2}$, $v_{3}$, $J$, symmetry). The generalized nomenclature used for the NOSD calculation, includes polyad number, symmetry, $J$, and eigenvalue ranking number (i.e., $P$, symmetry, $J$, $N$). When converting to the spectroscopic nomenclature, it is necessary to retain $N$ for uniqueness, since the other parameters are retained, or can be calculated (e.g., $P = 2V_{1}+V_{2}+4V_{3}$, where $V_{1}$, $V_{2}$, and $V_{3}$ are the quantum numbers of the normal modes of vibration). Hence, $N$ has been added to the global quantum numbers provided for $^{14}$N$_{2}$$^{16}$O. Five bands of $^{14}$N$_{2}$$^{16}$O in HITRAN2016 did not include symmetry for some of the quantum number assignments (i.e, 0000-0000, 1000-1000, 0200-0200, 0110-0000, 0600-1000), but were provided in \textit{NOSD$^{+}$}. For internal consistency, line assignments from \textit{NOSD$^{+}$} have been applied to all lines (included those added from HITRAN2016) to avoid accidental doubling of assignments.

Additional isotopologues contained in HITRAN2016 have also been added to \textit{NOSD$^{+}$}. One should keep in mind that the lack of bands for high temperature of these isotopologues are not sufficient for high-temperature applications, but are provided for consistency with HITRAN.

\subsection{Comparisons at 800~K}

A lack of high-temperature experimental observations of N$_{2}$O makes it difficult to validate the \textit{NOSD$^{+}$} line list at high temperatures. \citet{2016JQSRT.177...43T} provide a demonstration for the suitability of the \textit{NOSD-1000} line list for high temperatures, by comparing to a low-resolution measurement between 2025--2275~cm$^{-1}$ (FWHM = 1.0~cm$^{-1}$) at 873~K \citep{1997JQSRT..57..477R}. As expected, \textit{NOSD$^{+}$} performs equally well since all lines that appear at 1000~K with $S>1.0\times10^{-26}$ cm/molecule have been included. To compare the accuracy of this line list, it is also necessary to have a comparison at high resolution.

\begin{figure}[t!]
\centering
\includegraphics[scale=0.3, trim={0 0 0 0},clip]{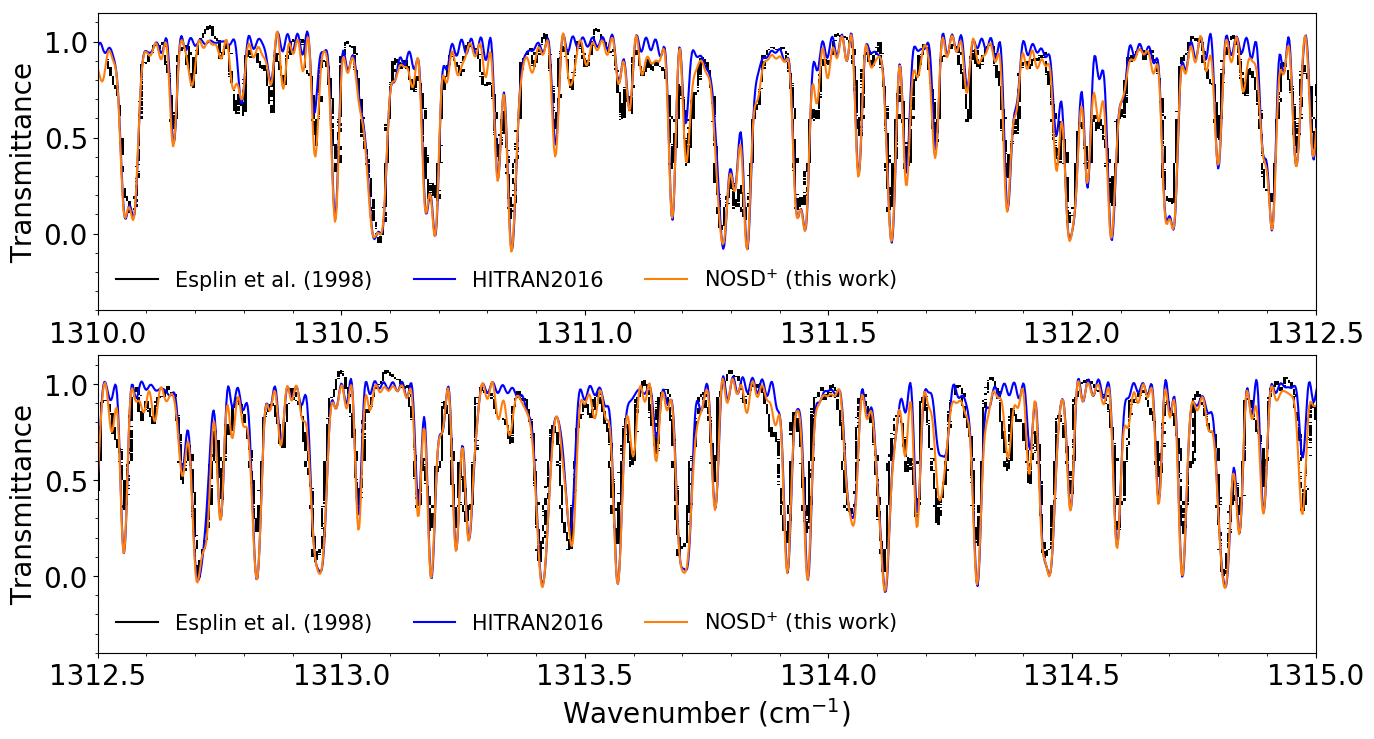}
\caption{Calculated transmission spectra of N$_{2}$O at 800~K, using line lists from HITRAN2016 (blue) and \textit{NOSD$^{+}$} (orange, and included in this work). \citet{1988AcMik...2..403E} data (in black) is taken as an overlaid image with an estimated transmittance scale. The spectra are calculated using HAPI \cite{2016JQSRT.177...15K} for a sample pressure of 4~Torr at 800~K and convolved to an FTS resolution of 0.02~cm$^{-1}$ (FWHM). \label{n2o_esplin}}
\end{figure}

\citet{1988AcMik...2..403E} measured the spectrum of N$_{2}$O between $1200 - 1360$ cm$^{-1}$ at a temperature of 800~K and pressure of 4~Torr. The data of \citet{1988AcMik...2..403E} were unavailable as supplementary files, so a spectral overlay (with an estimated transmittance scale) had to be used for comparisons. A limited high-resolution comparison can be made between $1310 - 1315$ cm$^{-1}$ in the vicinity of the $\nu_{1}$ fundamental. Figure~\ref{n2o_esplin} displays a comparison between the measurement overlay and transmission spectra calculated using the HITRAN2016 and \textit{NOSD$^{+}$} line lists. An excellent agreement is observed for \textit{NOSD$^{+}$}, with an improvement over HITRAN2016 throughout the spectral region (although HITRAN2016 performs well in this limited comparison). These improvements are due to a general increase in the `baseline' of the calculated \textit{NOSD$^{+}$} spectrum (a consequence of the significant increase in hot band transitions), as well as the inclusion of missing spectral features (e.g., at 1313.35~cm$^{-1}$), which are not included in HITRAN2016.

\subsection{Overview of the \textit{NOSD$^{+}$} line list added to the HITEMP database} \label{n2o_summary}

\begin{figure}[t!]
\centering
\includegraphics[scale=0.3, trim={0 0 0 0}, clip] {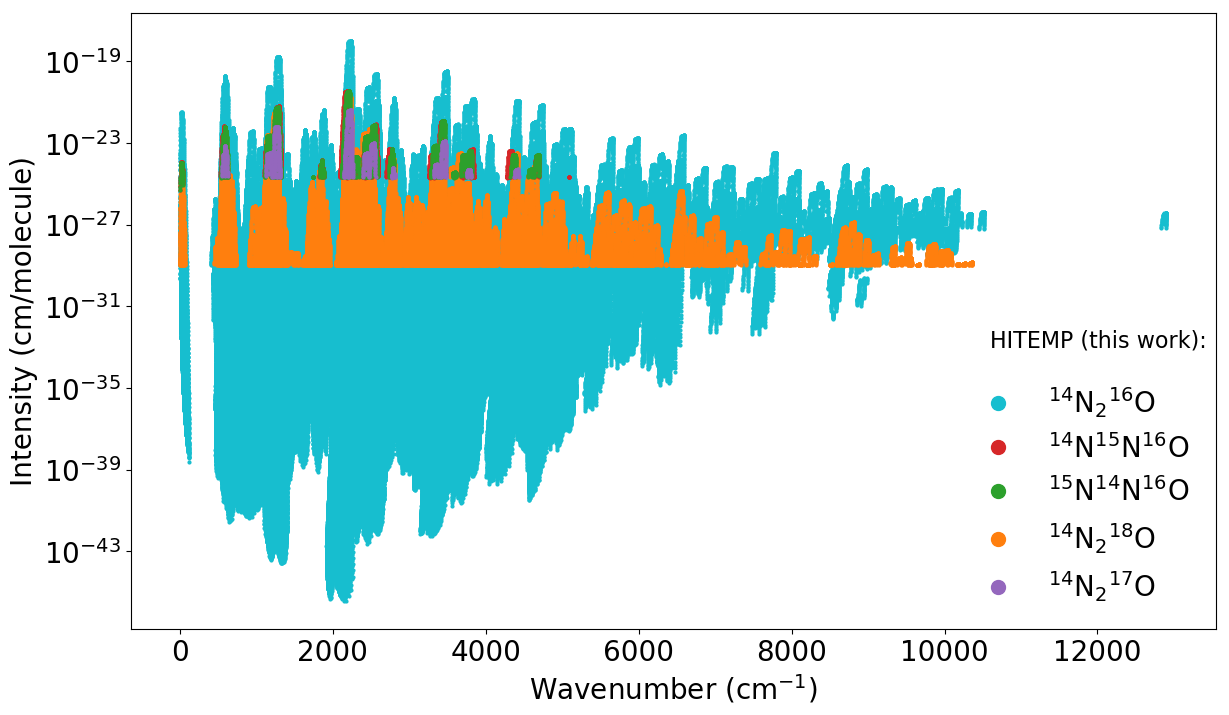}\vspace{0.2cm}
\caption{An overview of the \textit{NOSD$^{+}$} line list (this work) that is used to update the HITEMP database, with line intensities provided at 296~K. The isotopologues of N$_{2}$O have been separated and the corresponding number of lines is given in Table~\ref{tab_n2o_overview}. \label{n2o_full_overview}}
\end{figure}

N$_{2}$O was not previously included as part of HITEMP2010.  The \textit{NOSD$^{+}$} line list, produced for this work, provides the most complete N$_{2}$O line list currently available and has therefore been used to update HITEMP. The \textit{NOSD$^{+}$} line list is summarized in Table~\ref{tab_n2o_overview} and a full spectral overview is shown in Fig.~\ref{n2o_full_overview}.

\begin{table*}[t!]
\begin{center}
\caption{Summary of the N$_{2}$O line list used to update HITEMP for this work. \label{tab_n2o_overview} }
\begin{tabular}{lccccc}
\hline
 & & \\
Isotopologue               & $^{14}$N$_{2}$$^{16}$O  &  $^{14}$N$^{15}$N$^{16}$O  &  $^{15}$N$^{14}$N$^{16}$O  & $^{14}$N$_{2}$$^{18}$O  &  $^{14}$N$_{2}$$^{17}$O   \\
 & & \\
\hline
Isotopologue line count    &        3,499,212        &   4,222                    &   4,592                    &   116,694               &   1,705     \\
Natural abundance          &         0.990333        &   0.003641                 &   0.003641                 &   0.001986              &   0.000369  \\
$\nu_{min}$ (cm$^{-1}$)               &            0.838        &   5.028                    &   4.858                    &   0.791                 &   550.956   \\
$\nu_{max}$ (cm$^{-1}$)              &     12,899.137          &   5075.714                 &   4703.049                 &  10,363.675             &   4426.259  \\
$S_{min}$ (cm/molecule)                 &  3.750$\times10^{-46}$  &  5.220$\times10^{-26}$     &  4.720$\times10^{-26}$     &  1.000$\times10^{-29}$  &  2.016$\times10^{-26}$ \\
$S_{max}$ (cm/molecule)                 &  1.004$\times10^{-18}$  &  3.423$\times10^{-21}$     &  3.512$\times10^{-21}$     &  1.931$\times10^{-21}$  &  4.016$\times10^{-22}$  \\
\hline
\end{tabular}
\end{center}
\end{table*}

\section{Nitric Oxide (molecule 8)}\label{mol_no}

\subsection{Description of NO line lists}\label{line_lists}
There have been six line lists of NO that have been considered for inclusion to the HITEMP database. These line lists are briefly described below.

\subsubsection{NO in HITRAN2016}\label{list_hitran}
The NO line parameters in HITRAN2016 \cite{2017JQSRT.203....3G} include data for three isotopologues, specifically $^{14}$N$^{16}$O, $^{15}$N$^{16}$O and $^{14}$N$^{18}$O, with natural abundances of 0.993974, 0.003654 and 0.001993, respectively.

The $^{14}$N$^{16}$O data constitute 103,701 lines of the X~$^{2}\Pi_{\Omega'}$--X~$^{2}\Pi_{\Omega''}$ electronic transition including all $\Omega=1/2$ and $3/2$ components. The maximum vibrational level is $v_{max} = 14$ with a maximum $\Delta v = 5$ and $J_{max} = 125.5$ between $ 0 - 9273.214$ cm$^{-1}$. $\Lambda$-doubling is included for bands $v' - v'' = 0-0$, $1-1$, $1-0$, $2-1$, $2-0$ and $3-1$, with hyperfine splitting included where available. The majority of these data were provided by \citet{1998JQSRT..60..825G} and described in \citet{2005JQSRT..96..139R}. Some small corrections were made during the production of HITRAN2008 \cite{2009JQSRT.110..533R}, with the inclusion of 948 magnetic dipole transitions included for the  $0-0$ band. Recent analysis by \citet{SULAKSHINA2018171} observed small systematic shifts for unresolved $\Lambda$-doublets of $1/2 - 3/2$ sub-bands, corresponding to the additions from \citet{1998JQSRT..60..825G}.

The inclusion of the isotopologue data for NO is described in detail by \citet{2017JQSRT.203....3G}. For the $^{15}$N$^{16}$O and $^{14}$N$^{18}$O isotopologues, there are 699 and 679 lines of the $1-0$ fundamental band of X~$^{2}\Pi_{\Omega'} -$X~$^{2}\Pi_{\Omega''}$, respectively.

\subsubsection{NO in HITEMP2010}\label{list_hitemp}
The line parameters of the 115,610 lines of NO contained in HITEMP2010 \cite{2010JQSRT.111.2139R} include the three isotopologues also in HITRAN2016, and have remained unchanged since that publication. They primarily include data from \citet{1998JQSRT..60..825G} with $v_{max}' = 14$,  $\Delta v_{max} = 5$ and $J_{max} = 125.5$. An  intensity threshold was not applied, meaning some intensities are $<1.0\times 10^{-99}$ cm/molecule. HITRAN2016 updates for NO have not been applied to HITEMP2010, meaning there are small inconsistencies when compared to HITRAN2016. Consequently, the HITEMP2010 line list has not been used as part of  this update.

\subsubsection{NO in the CDMS}\label{list_cdms}
Recent FarIR/microwave data from the Cologne Database for Molecular Spectroscopy, CDMS \cite{2016JMoSp.327...95E}, have been made available by \citet{2015JMoSp.310...92M}. These data contain five isotopologues of NO, including $^{14}$N$^{16}$O, $^{14}$N$^{18}$O and $^{15}$N$^{16}$O (along with $^{14}$N$^{17}$O and $^{15}$N$^{18}$O that were not used as part of this update). A fit for all isotopolgues was performed using SPFIT \cite{1991JMoSp.148..371P}, and included hyperfine splitting. For each isotopologue, the $0-0$ pure rotational transitions have been determined using SPCAT \cite{1991JMoSp.148..371P}. For $^{14}$N$^{16}$O, the $1-1$ band was also included.

The data for the $0-0$ and $1-1$ bands of $^{14}$N$^{16}$O are available to download\footnote{\href{https://cdms.astro.uni-koeln.de/classic/entries/}{https://cdms.astro.uni-koeln.de/classic/entries/}} as two files (\texttt{c030517.cat} and \texttt{c030518.cat}) and contain 2955 and 1626 lines respectively, with $J_{max}=40.5$. These files have been converted from the standard CDMS format into that of HITRAN. The electric dipole selection rules (Q-branch: $e \rightarrow f$, $f \rightarrow e$; P-/R-branches: $e \rightarrow e$, $f \rightarrow f$) are used to identify and assign magnetic dipole transitions, since these selection rules are reversed. The line strengths have been adjusted for isotopic abundance (therefore multiplied by 0.993974) and differences between database reference temperatures ($T_{\textrm{\scriptsize{CDMS}}}=300$~K, $T_{\textrm{\scriptsize{HITRAN}}}=296$~K) have been accounted for (see Appendix of Ref. \cite{2009JQSRT.110..533R}). The total internal partition sums, TIPS2017 \citep{2017JQSRT.203...70G}, have been used to scale temperatures. The conversion from CDMS to HITRAN format is different for each $\Omega$ and depends on $J$, such that for $J \leq 5.5$ the CDMS quantum number code 0 refers to $\Omega={1/2}$ and the CDMS quantum number code 1 refers to $\Omega={3/2}$. For $J \geq 6.5$, this conversion is reversed. Magnetic dipole transitions are labelled with $m$ in the HITRAN notation, and these lines are included alongside electric dipole transitions ($P$-, $Q$-, $R$-branches), quadrupole transitions ($O$-, $S$-branches), and octupole transitions ($N$-, $T$-branches).

Wavenumber and intensity uncertainty codes for the CDMS line list are described in \ref{app_err_no}.

\begin{figure}[t!]
\centering
\includegraphics[scale=0.30, trim={0 0 0 0}, clip ]{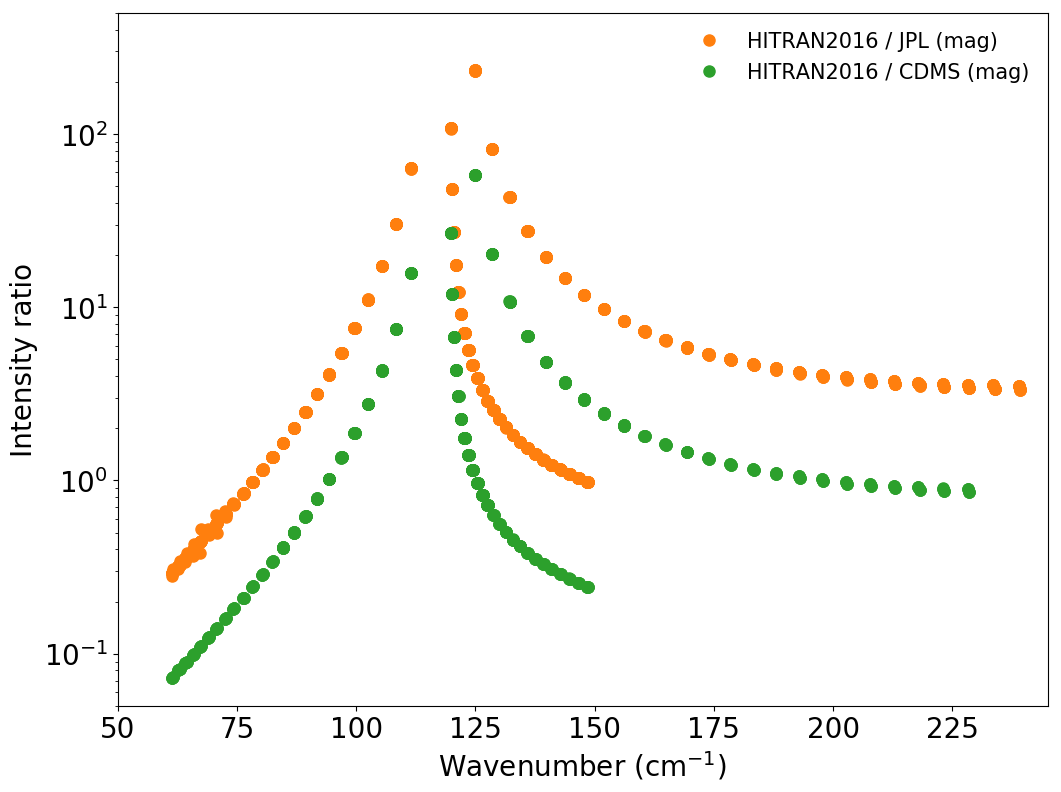}
\caption{Ratio of CDMS and JPL magnetic dipole transition intensities (compared to HITRAN2016) for the $\Omega =1/2 - 3/2$ component of the  $0-0$ band of $^{14}$N$^{16}$O.} \label{no_cdms}
\end{figure}

Magnetic dipole transitions in CDMS for the $\Omega = 1/2 - 3/2$ component of the $0-0$ band of $^{14}$N$^{16}$O have been compared to HITRAN2016 magnetic dipole transitions provided by \citet{1998JQSRT..60..825G}. These magnetic dipole transitions, in the  $50 - 250$ cm$^{-1}$ range, are stronger than the ``forbidden'' electric dipole transitions for the same band. CDMS and HITRAN2016 show similar agreement when compared to experimental measurements of the $Q$-branch (near 124~cm$^{-1}$) \cite{1999JMoSp.196....5V}. However, significant disagreement of almost two orders of magnitude can be seen for other transitions of this band (Fig.~\ref{no_cdms}). This disagreement is likely due to the alternative approaches used to the calculate the  magnetic transition dipole moment. The values from \citet{1998JQSRT..60..825G} have previously been used in atmospheric retrievals and it was decided to retain these values for this work. Hence, no magnetic dipole transitions from the CDMS database have been included as part of this update. These comparisons do highlight a need for additional work to validate these data.

The same method was used to convert the $^{15}$N$^{16}$O and $^{14}$N$^{18}$O isotopologues from CDMS to HITRAN format. There are 1626 lines of $^{15}$N$^{16}$O (file \texttt{c031512.cat}) and 3027 lines of $^{14}$N$^{18}$O (file \texttt{c032513.cat}). The magnetic dipole transitions have also been excluded from these lists.

\subsubsection{NO in the JPL catalogue}\label{list_jpl}
In addition to the CDMS data, an extended $0-0$ line list for $^{14}$N$^{16}$O is available to download\footnote{\href{https://spec.jpl.nasa.gov/ftp/pub/catalog/catdir.html}{https://spec.jpl.nasa.gov/ftp/pub/catalog/catdir.html}} through the Jet Propulsion Laboratory (JPL) Molecular Spectroscopy Catalog \cite{1998JQSRT..60..883P}. This line list (file \texttt{c030008.cat}) contains 9765 lines and is calculated using the same programs as the CDMS line lists (i.e., SPFIT/SPCAT), but with alternative fitting coefficients and a $J_{max}=71.5$ \cite{2010..web.Drouin}.

\begin{figure}[t!]
\centering
\includegraphics[scale=0.3, trim={0 0 0 0}, clip] {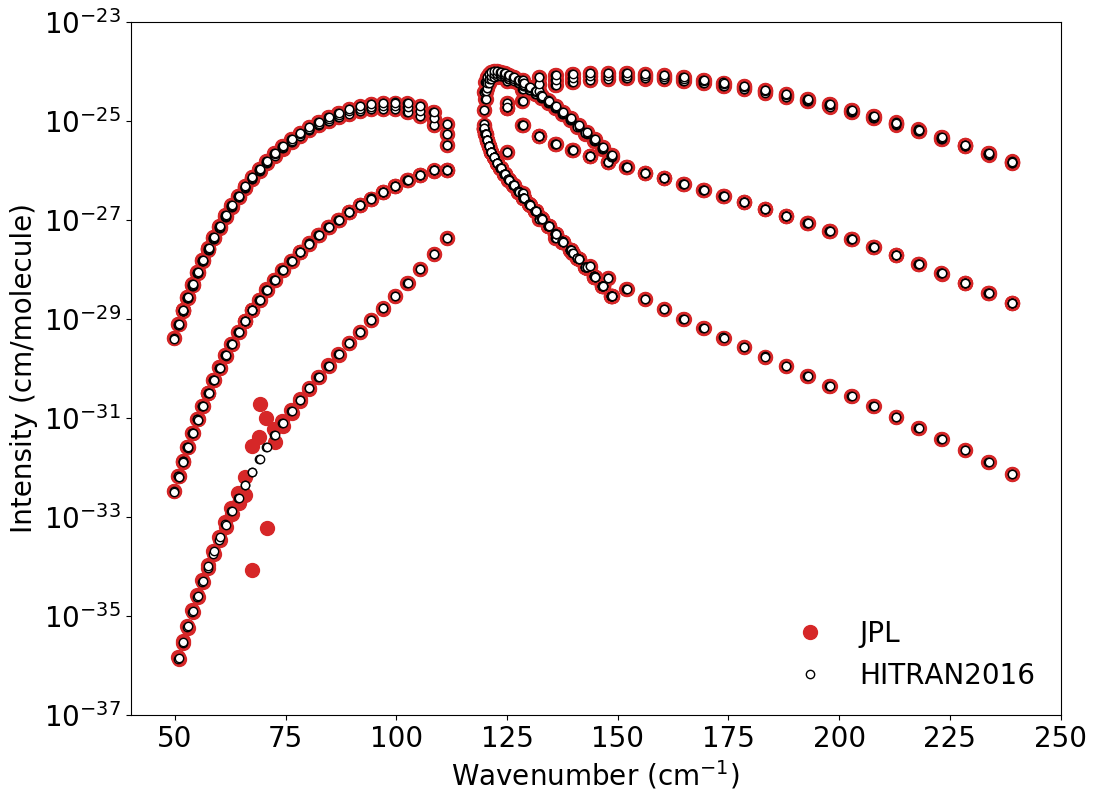}
\caption{Comparison of matched line intensities for the $0-0$ band ($\Omega'=3/2$, $\Omega'=1/2$) between the JPL and HITRAN2016 line lists.\label{no_jpl}}
\end{figure}

The JPL data set was converted into the HITRAN format following the same procedure as described for the CDMS data (also using TIPS2017).Magnetic dipole transitions from the JPL line list were also excluded, as they were seen to exhibit a greater disagreement than the CDMS line list when compared to HITRAN2016 values (Fig.~\ref{no_cdms}). A comparison of the $0-0$ band ($\Omega'=3/2$, $\Omega'=1/2$) to HITRAN2016 indicates large intensity discontinuities near 70 cm$^{-1}$, as shown in Fig.~\ref{no_jpl}. These discrepancies occur for the $P$-branch where $F' = F''+1$. The JPL lines therefore do not improve on the current HITRAN2016 data and have been excluded from this update.

\subsubsection{ExoMol: \textit{NOname}}\label{list_noname}

A comprehensive semi-empirical line list, \textit{NOname}, has been published by \citet{2017MNRAS.470..882W} for six isotopologues of NO, as part of the ExoMol project \citep{2016JMoSp.327...73T}. These data were earmarked by \citet{2017JQSRT.203....3G} to be included as part of the next HITRAN/HITEMP update. The line lists are calculated for six isotopologues: $^{14}$N$^{16}$O, $^{15}$N$^{16}$O, $^{14}$N$^{18}$O, which are included in HITRAN2016, as well as $^{14}$N$^{17}$O, $^{15}$N$^{17}$O, $^{15}$N$^{18}$O that have not been included in this update. Each line list is constructed using empirical energy levels (and line positions) and high-level \textit{ab initio} intensities. The effective Hamilton is obtained from a fit to available experimental positions using SPCAT and SPFIT. Data from the CDMS line list \cite{2015JMoSp.310...92M} is also included. Secondly, a variational model is built by fitting to the experimentally obtained energy levels and positions using the \texttt{Duo} program for diatomic molecules \cite{2016CoPhC.202..262Y}. This enables the \textit{NOname} line list to be extended to include higher  vibration bands and rotational levels. The wavenumber coverage is 0--40,040 cm$^{-1}$, with $J_{max}=184.5$ and $v'_{max}=51$ (with $\Delta v_{max}=44$), and the associated partition function ($Q_{\textrm{\scriptsize{ExoMol}}}$) is applicable up to 5000~K. The full \textit{NOname} line list is constructed by stitching together lines from both methods at $J=99.5$ and $v=28$ (for $^{14}$N$^{16}$O). Hence, line positions and intensities for $J \geq 100.5$ and $v \geq 29$ are taken from \texttt{Duo}.

The data for each isoptopologue are divided into the standard `states' and `trans' ExoMol files\footnote{\href{http://exomol.com/data/molecules/NO/14N-16O/NOname/}{http://exomol.com/data/molecules/NO/14N-16O/NOname/}}, which separates the state assignments from the individual transitions. Each transition has been converted into the HITRAN format (i.e., at reference temperature 296~K) with the appropriate transition information included \cite{2017JQSRT.203....3G}. The intensities have been adjusted for natural abundance and the partition function from TIPS2017 \citep{2017JQSRT.203...70G} has been used to remain consistent with current HITRAN2016 data (and the line lists described above). Wavenumber and intensity uncertainty codes are described in \ref{app_err_no}.

The $^{15}$N$^{16}$O and $^{14}$N$^{18}$O isotopologues show a slight intensity decrease when compared to HITRAN2016. However, \citet{Heinrich2009} have observed that HITRAN2016 intensities of $^{15}$N$^{16}$O appear to be overestimated when monitoring the isotopic ratios of NO in human breath. Combined with the good overall agreement observed for $^{14}$N$^{16}$O, the $^{15}$N$^{16}$O and $^{14}$N$^{18}$O \textit{NOname} line lists improve on the HITRAN2016 values and have also been included as part of this update.

\subsection{Comparisons of the \textit{NOname} line list}\label{comp_noname}

\begin{figure}[t!]
\centering
\includegraphics[scale=0.3, trim={0 0 0 0}, clip] {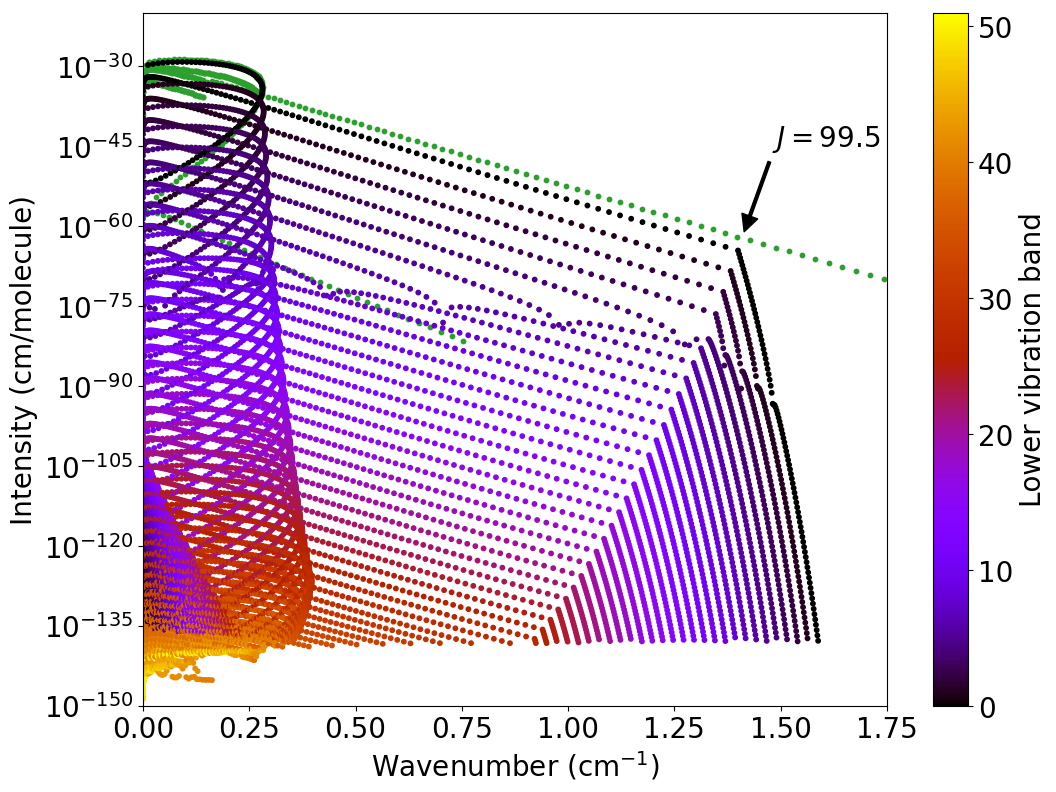}
\caption{Intensities of $^{14}$N$^{16}$O lines contained in the \textit{NOname} line list, with the color referring to the lower vibrational level ($v''$). The stitching point is clearly visible for vibrational bands at $J = 99.5$, as indicated by the arrow. For reference, the HITRAN2016 values for the $0-0$ band have been plotted in green. These HITRAN values are approximately 2 orders of magnitude stronger than the corresponding values in the \textit{NOname} line list. \label{no_stitch}}
\end{figure}

The stitching point used in \textit{NOname} leads to an unexpected line position error for lines with $J > 99.5$ as shown in Fig.~\ref{no_stitch}. The line position errors are most noticeable for the $Q$-branch with $\Delta\Omega = 0$ of the $\Delta v = 0$ bands and arise from  the change in energy levels from the ones calculated with SPFIT/SPCAT to those calculated with \texttt{Duo}. Both methods have their disadvantages when extrapolating far beyond experimental observations. However, \texttt{Duo} energy levels are steadily declining from the real values even within the the range probed by experiments. This especially concerns $\Lambda$-doubled splitting which is not correctly modeled by \texttt{Duo} even at low $J$'s. Here the differences lead to a discontinuity at $J = 99.5$ for all vibrational bands. This effect is seen for all bands throughout the \textit{NOname} line list. As a consequence of this discontinuity, all lines with $J > 99.5$ and $v' > 29$ have been excluded from this $^{14}$N$^{16}$O update. For the $^{15}$N$^{16}$O and $^{14}$N$^{18}$O isotopologues, the \texttt{Duo} stitching point occurs at a lower vibrational level; therefore transitions for these isotopologues with $v' > 9$ have been excluded.

For $\Delta v = 0$, the $P$-, $Q$- and $R$-branches of the $\Omega'-\Omega'' = 1/2 - 3/2$ components show good agreement when compared to HITRAN2016. However, the $Q$-branches with $\Delta\Omega = 0$ of the $\Delta v = 0$ band display significant intensity differences when compared to values in HITRAN2016 (up to 3 orders of magnitude) and have also been excluded from this update. Indeed, this difference can be seen in Fig.~\ref{no_stitch}, where values from HITRAN2016 are seen to be significantly stronger then the corresponding values in \textit{NOname}. These transitions are between the $\Lambda$-doubling components and the intensities calculated with \texttt{Duo} are not reliable here due to the aforementioned problem with modelling of the energy splitting associated with this doubling. Interestingly this effect does not influence intensities in other bands as the $\Lambda$-doubling splitting is much smaller than separation between rotational and especially rovibrational levels.  

The MARVEL approach \cite{2007JMoSp.245..115F, 2012JQSRT.113..929F} used to obtain empirical energy levels in the calculation of \textit{NOname} is similar to the RITZ method \cite{2003JQSRT..82..165T} used by \citet{SULAKSHINA2018171}. Their comparisons between energy levels for resolved $\Lambda$-doublets of $^{14}$N$^{16}$O coincide within the experimental uncertainty below $J=37.5$.

An additional $\Delta v \leq 16$ filter was applied because of numerical artifacts (e.g., noise) observed in the variational transition dipole moments for $\Delta v > 16$ (see Fig.~4 of Ref. \cite{2017MNRAS.470..882W}).  The nature of these artifacts, and their impact on the calculated intensities of diatomic molecules, has previously been described in detail by \citet{Medvedev2015,Medvedev2016}. These data filters were also applied to the $^{15}$N$^{16}$O and $^{14}$N$^{18}$O line lists of \textit{NOname}.

\begin{figure}[t!]
\centering
\includegraphics[scale=0.3, trim={0 0 0 0}, clip] {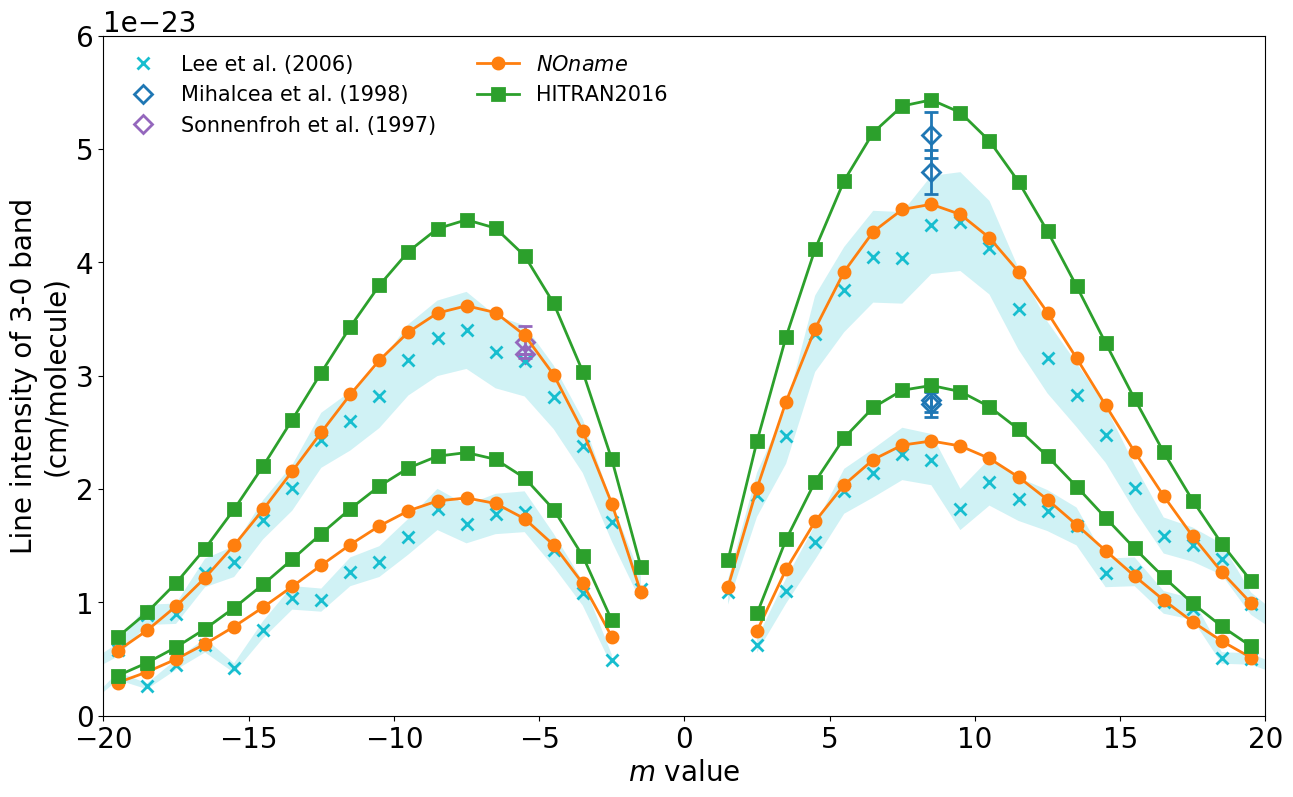}
\caption{Observed experimental intensities for the $3-0$ band of $^{14}$N$^{16}$O \cite{2006InPhT..47..227L, 1998MeScT...9..327M, 1997ApOpt..36.7970S}, compared to line lists of HITRAN2016 and \textit{NOname} \cite{2017MNRAS.470..882W}. The $P$- and $R$-branches are shown with $\Omega'' = 1/2$ components stronger that $\Omega'' = 3/2$ for each branch. Error bars of $\pm10$\% fron \citet{2006InPhT..47..227L} are shown as shaded areas. The  \textit{NOname} lines for this band have been incorporated into the HITEMP line list for this work. \label{no_30band}}
\end{figure}

\begin{figure}[t!]
\centering
\includegraphics[scale=0.3, trim={0 0 0 0}, clip] {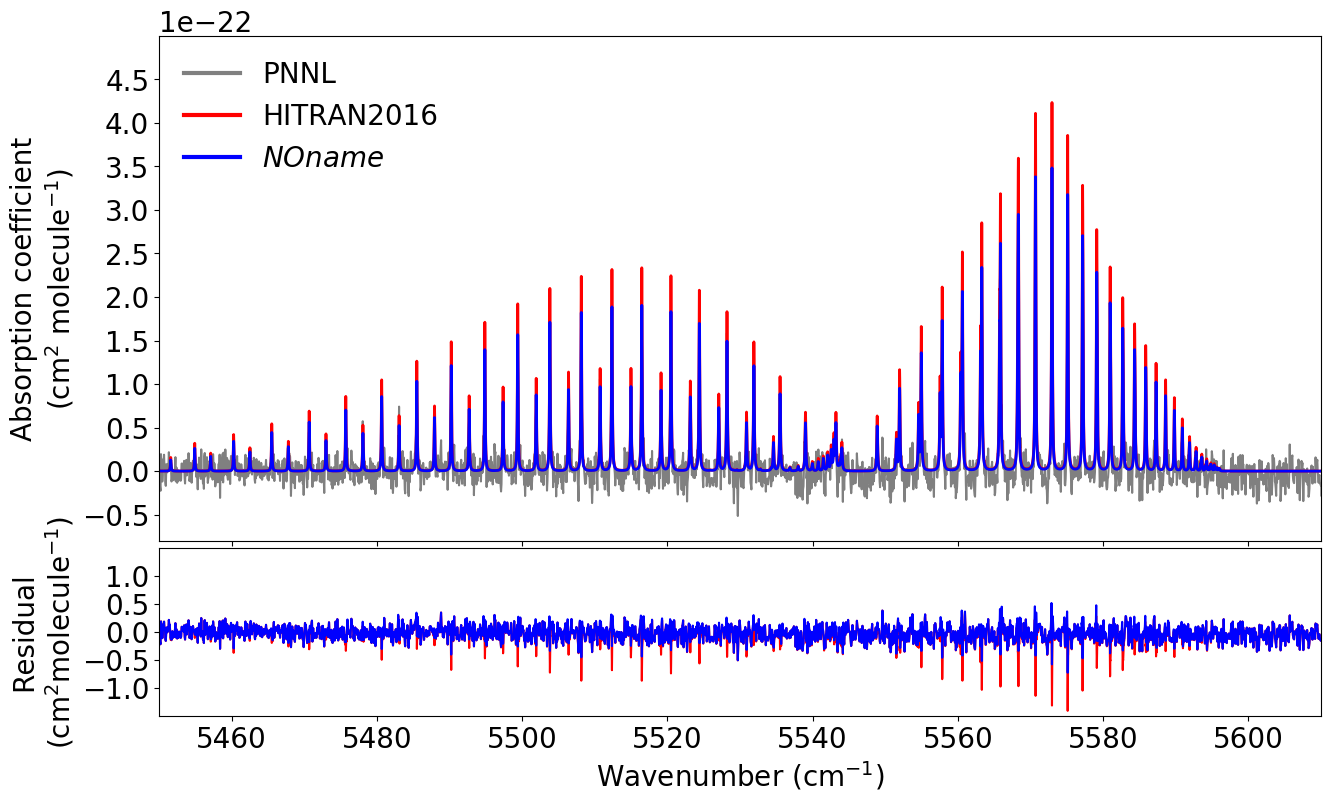}
\caption{Experimental absorption cross sections for the $3-0$ band of $^{14}$N$^{16}$O at 25$^{\circ}$C from PNNL \cite{2004ApSpe..58.1452S} compared to calculated absorption cross sections using \textit{NOname} \cite{2017MNRAS.470..882W} and HITRAN2016. The  \textit{NOname} lines for this band have been incorporated into the HITEMP line list for this work. HAPI \cite{2016JQSRT.177...15K} has been used for the calculations. \label{no_30band_pnnl}}
\end{figure}

Furthermore, the \textit{NOname} line intensities have been compared to experimental observations of overtone bands. \citet{2006InPhT..47..227L} measured line intensities for the $2-0$, $3-0$, $4-0$, $5-0$, and $6-0$ bands, and \citet{2006JChPh.124h4311B} measured the $7-0$ band. For the $3-0$ band, the \textit{NOname} intensities agreed well with experiment \cite{2006InPhT..47..227L}, but were approximately 15\% weaker than HITRAN2016 (Fig.~\ref{no_30band}). Diode-laser absorption measurements \cite{1998MeScT...9..327M, 1997ApOpt..36.7970S} were able to verify the \textit{NOname} line intensities for the $3-0$ band. A further comparison has been performed against experimental cross-sections from PNNL \cite{2004ApSpe..58.1452S}, and is shown in Fig.~\ref{no_30band_pnnl}. An absorption cross section calculated using HITRAN2016 displays a larger residual than the corresponding cross sections calculated using \textit{NOname}.


\begin{figure}[t!]
\centering
\includegraphics[scale=0.25, trim={0 0 0 0}, clip] {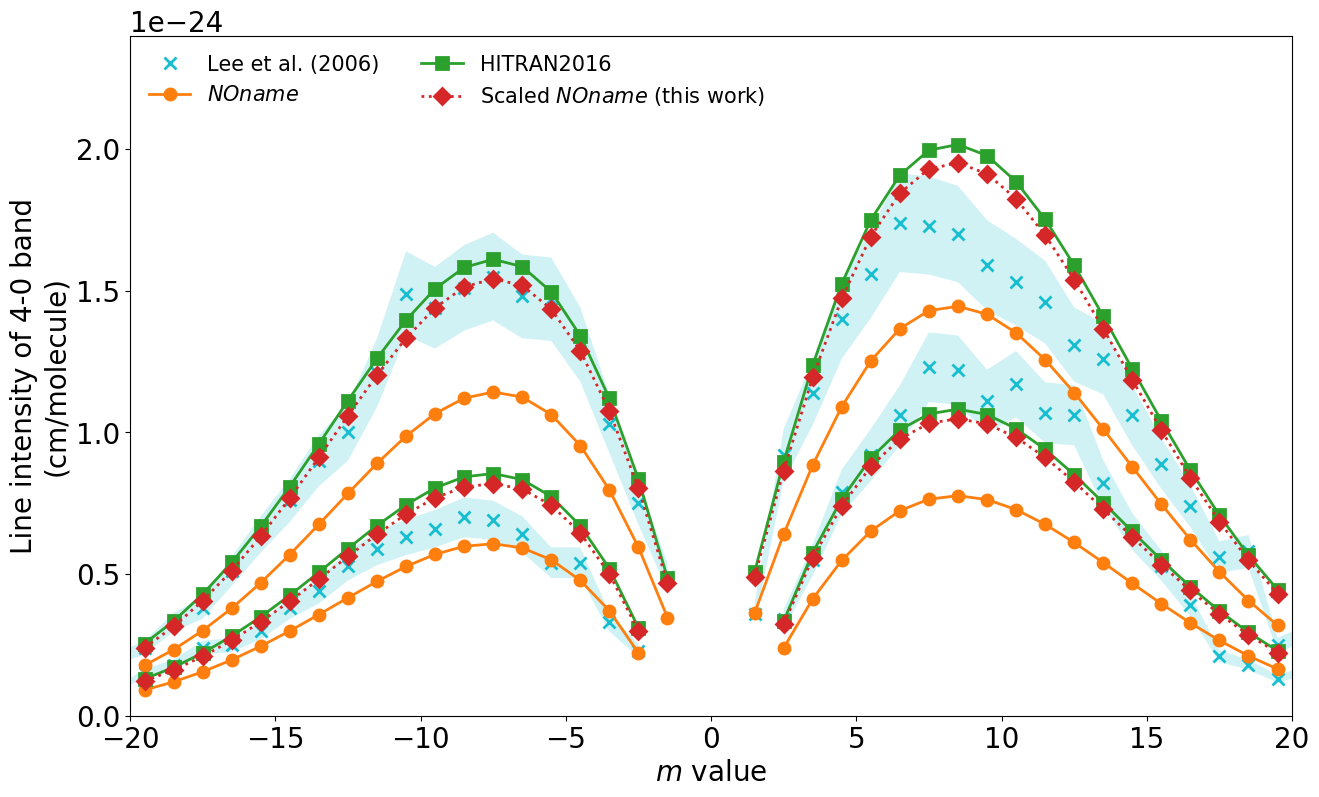}\\
\includegraphics[scale=0.25, trim={0 0 0 0}, clip] {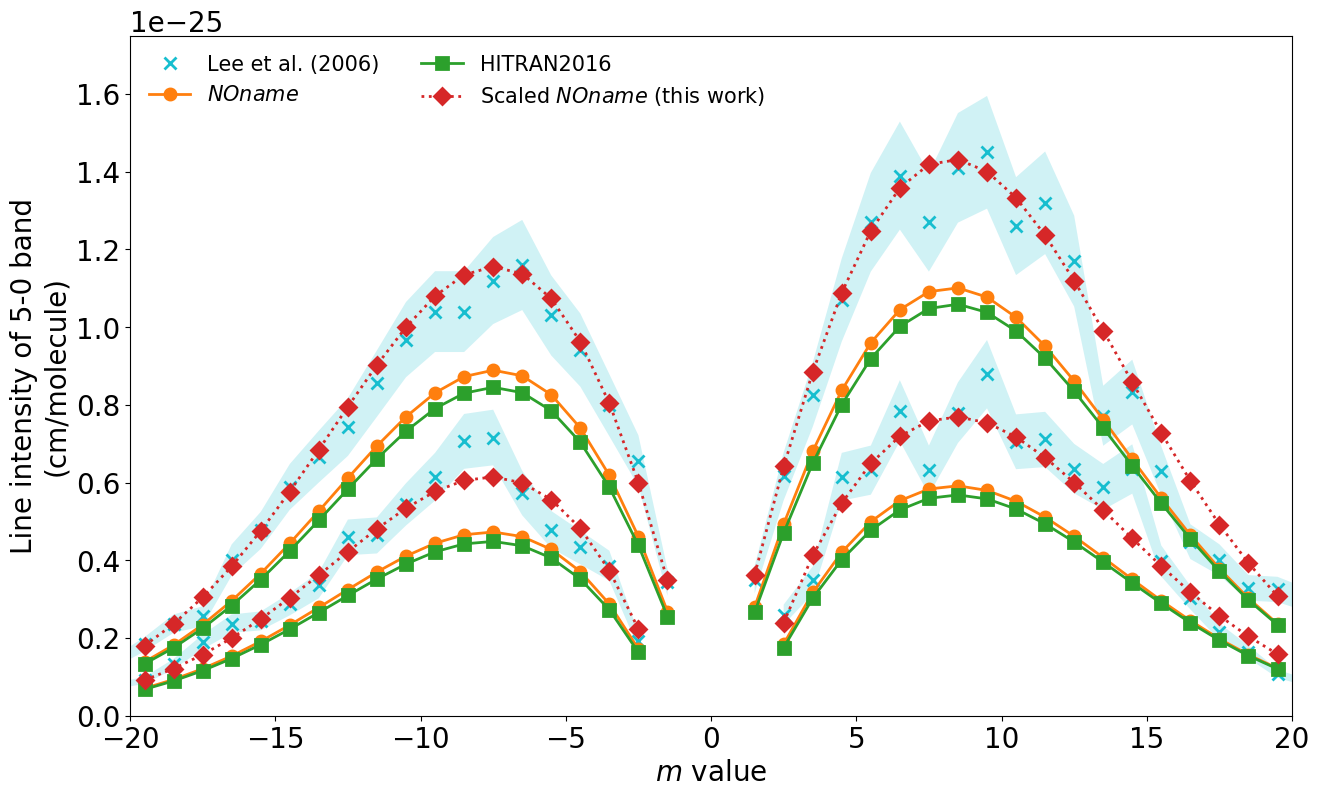}\\
\includegraphics[scale=0.25, trim={0 0 0 0}, clip] {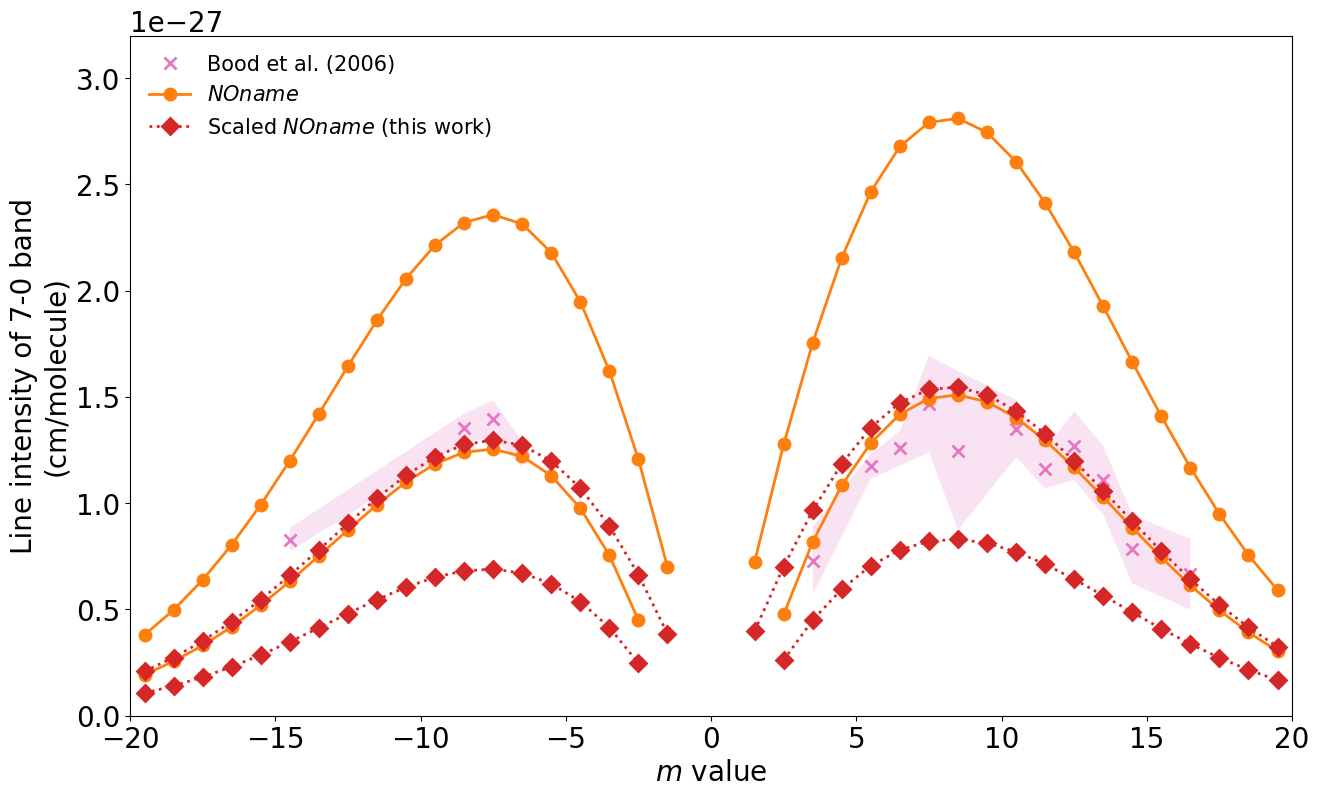}
\caption{Observed experimental intensities for the $4-0$ (top panel), $5-0$ (middle panel) and $7-0$ (lower panel) bands of $^{14}$N$^{16}$O \cite{2006InPhT..47..227L, 2006JChPh.124h4311B} compared to line lists of \textit{NOname} \cite{2017MNRAS.470..882W} and HITRAN2016.  The $P$- and $R$-branches are shown with $\Omega'' = 1/2$ components stronger that $\Omega'' = 3/2$ for each branch. $\Lambda$-doubled components have been added together for comparisons with experiment.  Error bars for experimental measurements are shown as shaded areas, with $\pm10$\% for \citet{2006InPhT..47..227L} and absolute values for \citet{2006JChPh.124h4311B}. Scaled \textit{NOname} intensities of 1.35, 1.30 and 0.55 for the $4-0$, $5-0$ and $7-0$ bands, respectively are also shown and have been incorporated into the HITEMP lie list for this work. \label{no_457}}
\end{figure}

Excellent agreement was observed for the majority of vibrational bands.  However some differences have been observed: these are shown in  Fig. \ref{no_457}. For the $4-0$ band, the \textit{NOname} intensities are too weak when compared to both HITRAN2016 and experiment \cite{2006InPhT..47..227L}. For the $5-0$ band, the \textit{NOname} line intensities show good agreement with HITRAN2016, but differ from experimental values \cite{2006InPhT..47..227L}. For the $7-0$ band, the difference between the \textit{NOname} line intensities and \citet{2006JChPh.124h4311B} is approximately a factor of two. Hence, a scale factor of 1.35, 1.30 and 0.55 has been applied to all $\Delta v=4$, $\Delta v=5$ and $\Delta v=7$ bands, respectively. The scaled \textit{NOname} intensities are also included in Fig. \ref{no_457} for comparison.

\subsubsection{HSPM2018}\label{list_hspm2018}

For atmospheric applications, it is important to include high rotational levels for NLTE observations.  However Section~\ref{comp_noname} describes the removal of these lines from \textit{NOname}. Therefore, for this work the SPCAT/SPFIT calculation used as part of \citet{2017MNRAS.470..882W} for $^{14}$N$^{16}$O has been extended to produce the HSPM2018 line list. This line list extends to $v_{max}'=29$ for $\Delta v=0$ and $v_{max}'=10$ for $\Delta v=1$, with $J_{max} = 184.5$. Hyperfine splitting has been ignored. The associated \textit{NOname} line intensities have been combined with these line positions to retain consistency. See \ref{app_err_no} for details regarding uncertainty codes.

\subsection{Combining line lists of NO}\label{list_combined}

This work describes the construction of a suitable NO line list to update HITEMP. The vast majority of lines are included from \textit{NOname}, with a hierarchical merging of additional line lists outlined below.

The reduced $^{14}$N$^{16}$O \textit{NOname} line list (described in Sect.~\ref{comp_noname}) was taken as the starting point. It extends up to $26,776$ cm$^{-1}$, with $\Delta v \leq 16$,  $v \leq 29$ and $J \leq 99.5$. Some of these bands have had their intensity scaled ($\Delta v=4$, $\Delta v=5$ and $\Delta v=7$). One should also note that only a small fraction of the $\Delta v=0$ band from the \textit{NOname} line list has been taken, for the reasons explained earlier. 

For the $0-0$ and $1-1$  bands, line positions and intensities from the CDMS line list (see Sect.~\ref{list_cdms}) have replaced any corresponding lines in \textit{NOname}. Lines from HITRAN2016 that make up the $0-0$, $1-1$, $1-0$, $2-1$, $2-0$ and $3-1$ bands have also replaced any corresponding lines in \textit{NOname}, providing they were not included in the CDMS line list (or from \citet{1998JQSRT..60..825G}). These bands retain their hyperfine splittings, and magnetic dipole transitions from HITRAN2016 for the $0-0$ band have also been included.

For bands with $\Delta v=0$ and $\Delta v=1$, lines from the HSPM2018 line list have been used to replace the lines from \textit{NOname} that were excluded because of the \texttt{Duo} stitching point. This includes $99.5 < J \leq 184.5$ and $v \leq 29$ for $\Delta v=0$, and  $99.5 < J \leq 184.5$ and $v \leq 10$ for $\Delta v=1$. For bands with $\Delta v=2$ to $\Delta v=5$, the line positions from \citet{1998JQSRT..60..825G} (and also in HITRAN2016)  with $99.5 < J \leq 124.5$ have also been used to replace the excluded lines from \textit{NOname}.

It should be noted that for $R$-branches of the $\Omega = 1/2 - 1/2$ and $\Omega = 3/2 - 3/2$ components of the  $0-0$ and $1-1$ bands, the CDMS line strengths differ by almost 20\% when compared to HITRAN2016 values. It is difficult to assess the validity of these intensity differences because the methods used for calculation are very different, and no experimental measurements exist to access intensities where the largest differences are observed. A slight intensity discontinuity will therefore be seen in the line list produced for this work.

\begin{figure}[t!]
\centering
\includegraphics[scale=0.3, trim={0 0 0 0}, clip] {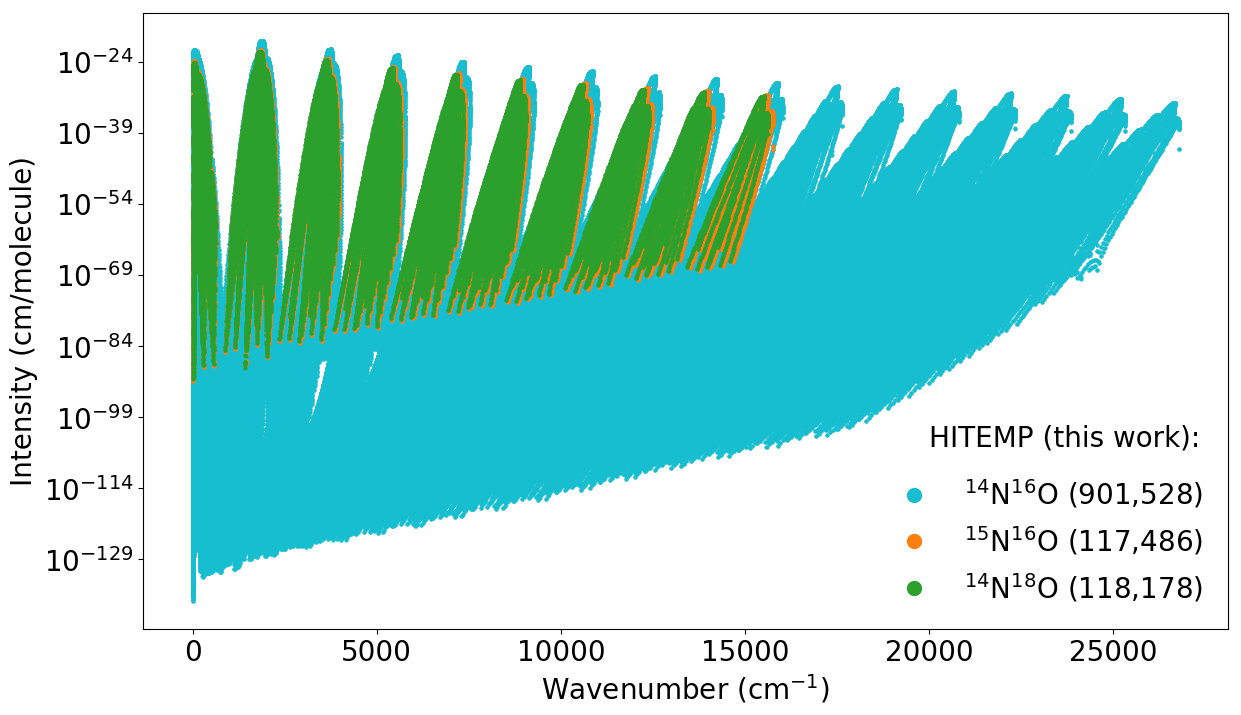}
\caption{An overview of the intensity and wavenumber coverage for the combined NO line lists produced for this work. Each isotopologue has been indicated, with the corresponding number of lines given in parenthesis.  The line intensities are provided at 296~K.\label{no_full_overview}}
\end{figure}

For the $^{15}$N$^{16}$O and $^{14}$N$^{18}$O isotopologues, only lines contained in the CDMS line lists have been used to replace those from \textit{NOname}. The combined line lists containing all three isotopologues are shown in Fig.~\ref{no_full_overview}.

\begin{figure}[t!]
\centering
\includegraphics[scale=0.3, trim={0 0 0 0}, clip] {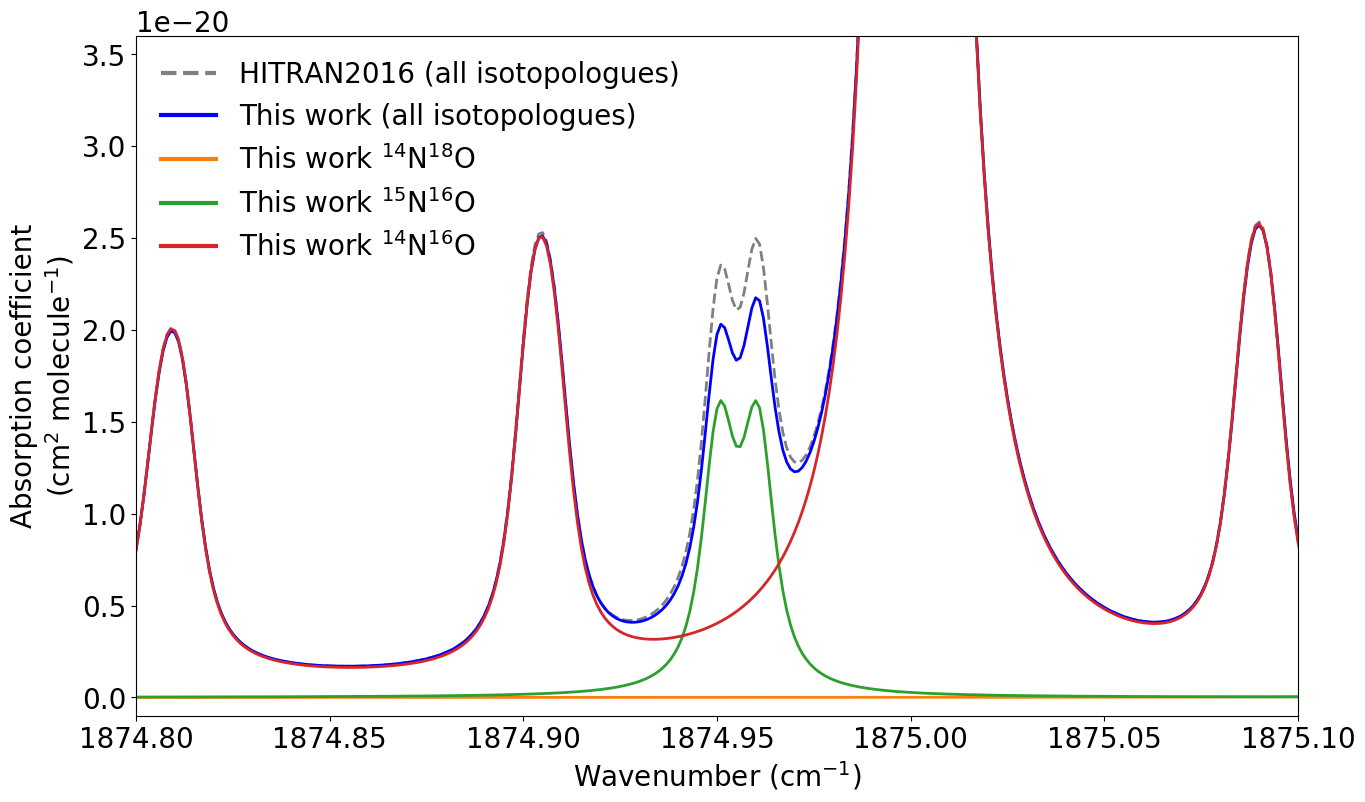}
\caption{A comparison between calculated absorption cross sections of NO at 296~K using HITRAN2016 and the line list produced for this work. The spectral region is the same as that used to monitor isotopic ratios of NO in human breath \cite{Heinrich2009}. A 16\% change in the contribution of $^{15}$N$^{16}$O is seen near 1875~cm$^{-1}$. All spectra have been calculated using HAPI \cite{2016JQSRT.177...15K}.\label{no_full_15no_exp}}
\end{figure}

\citet{Heinrich2009} have previously reported that the intensities of $^{15}$N$^{16}$O were overestimated in HITRAN. Their method uses a carbon monoxide laser to spectroscopically analyze human breath for isotopic ratios of NO near 1875~cm$^{-1}$ when monitoring a variety of health conditions. Fig.~\ref{no_full_15no_exp} shows the spectral region used by \citet{Heinrich2009} to measure the abundance of $^{15}$N$^{16}$O. The line list produced for this work shows a reduction in intensity for the  $^{15}$N$^{16}$O lines by approximately 16\% (when compared to HITRAN2016).  This supports the conclusion of \citet{Heinrich2009}, and would correct their retrieved abundances to within experimental uncertainty.

\subsection{Reanalysis of HITRAN2016 broadening and shift parameters for NO}\label{new_params}

HITRAN2016 contains inconsistent broadening parameters for NO caused by intermediate updates. This work provides an opportunity to reanalyze the available line broadening and pressure shift parameters to create a consistent line list.

\subsubsection{Pressure shifts for NO}\label{new_shifts}

Measurements of the pressure shifts were obtained from \citet{1994JMoSp.165..506S} for the fundamental vibration band. Calculating pressure shifts for additional vibration bands ($\delta_{v'-v''} (m)$) has been described by \citet{2000JChPh.113.9000H} using 
\begin{equation}\label{eqn_shifts}
  \delta_{v'-v''} (m) = \left[ \left( \frac{ v'-v'' }{ v_{r}'-v_{r}'' }\right) \delta_{v_{r}'-v_{r}''}^{+}(m)\right] + \delta_{v_{r}'-v_{r}''}^{-}(m) \; ,
\end{equation}
where $v_{r}'-v_{r}''$ refers to the reference band measurements, and $m=-J''$ for $P$-branch and $m=J''+1$ for $R$-branch transitions. In this case, $v_{r}'-v_{r}'' = 1-0$ which simplifies Eq.~\ref{eqn_shifts} to
\begin{equation}\label{eqn_shifts_b}
  \delta_{v'-v''} (m) = ( v'-v'')\delta_{1-0}^{+}(m) + \delta_{1-0}^{-}(m) \; ,
\end{equation}
\noindent
where the symmetric and asymmetric components are given by
\begin{equation}\label{eqn_shifts_sym}
 \delta_{1-0}^{+}(m) = \frac{1}{2} [ \delta_{1-0}(m) + \delta_{1-0}(-m) ] \; ,\; \textrm{and}
\end{equation}
\begin{equation}\label{eqn_shifts_asym}
 \delta_{1-0}^{-}(m) = \frac{1}{2} [ \delta_{1-0}(m) - \delta_{1-0}(-m) ] \; .
\end{equation}
\noindent
This method has previously been used for CO pressure shifts in HITRAN2016 \citep{2015ApJS..216...15L}.

\begin{figure}[t!]
\centering
\includegraphics[scale=0.3, trim={0 0 0 0}, clip] {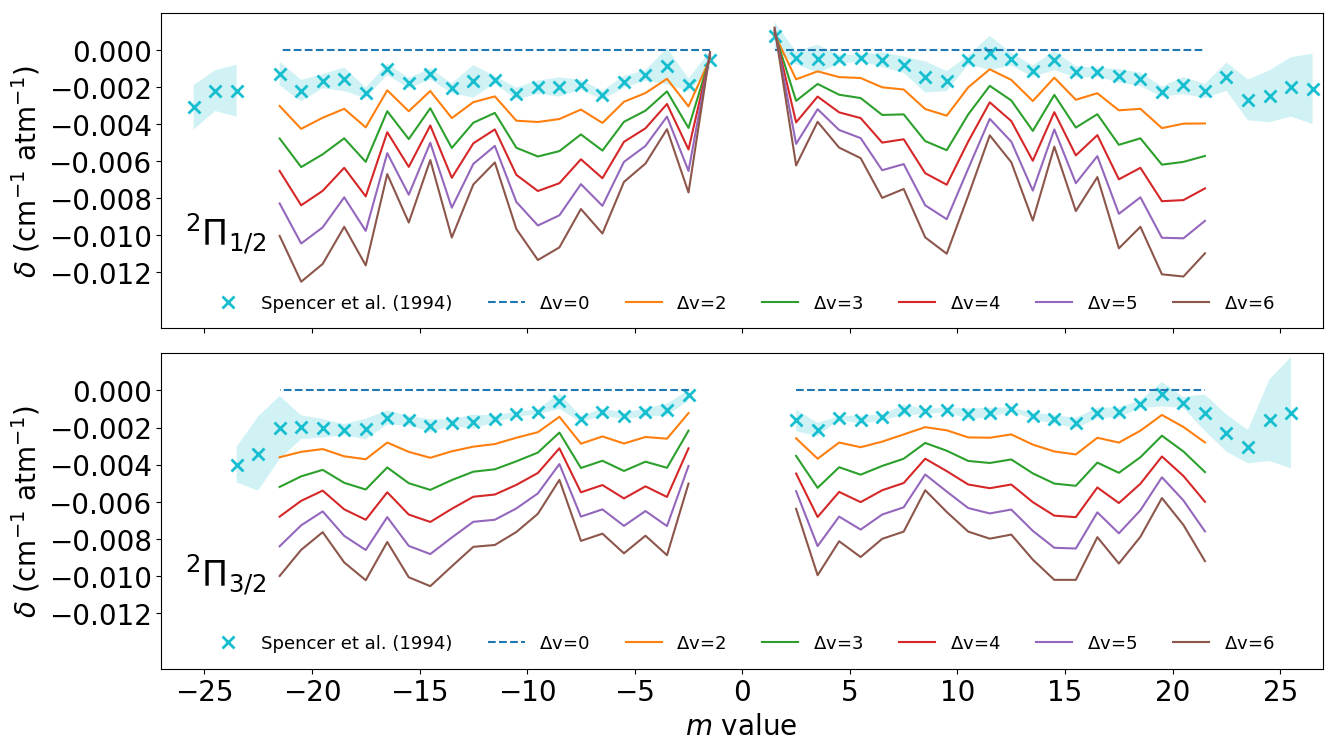}
\caption{Measured NO pressure shifts for the  $1-0$ band \cite{1994JMoSp.165..506S}, separated by $\Omega$ component. These measurements have been extended to additional vibrational bands for this work (using Eq.~\ref{eqn_shifts_b}), and are shown for for five higher overtones (up to $\Delta v=6$). A zero shift is applied for all  $\Delta v=0$ bands and is also plotted. \label{no_shifts}}
\end{figure}

Fig.~\ref{no_shifts} displays the extension of the pressure shifts for $\Delta v=0$ to 6. Further details of the applied pressures shifts is given in \ref{app_pressure_no}. This work marks a significant improvement on the previous average shift values for NO in HITRAN2016, which were only applied to the $1-0$ and $2-0$ bands.

\subsubsection{Broadening parameters for NO}\label{new_broad}

\begin{figure}[t!]
\centering
\includegraphics[scale=0.3, trim={0 0 0 0}, clip] {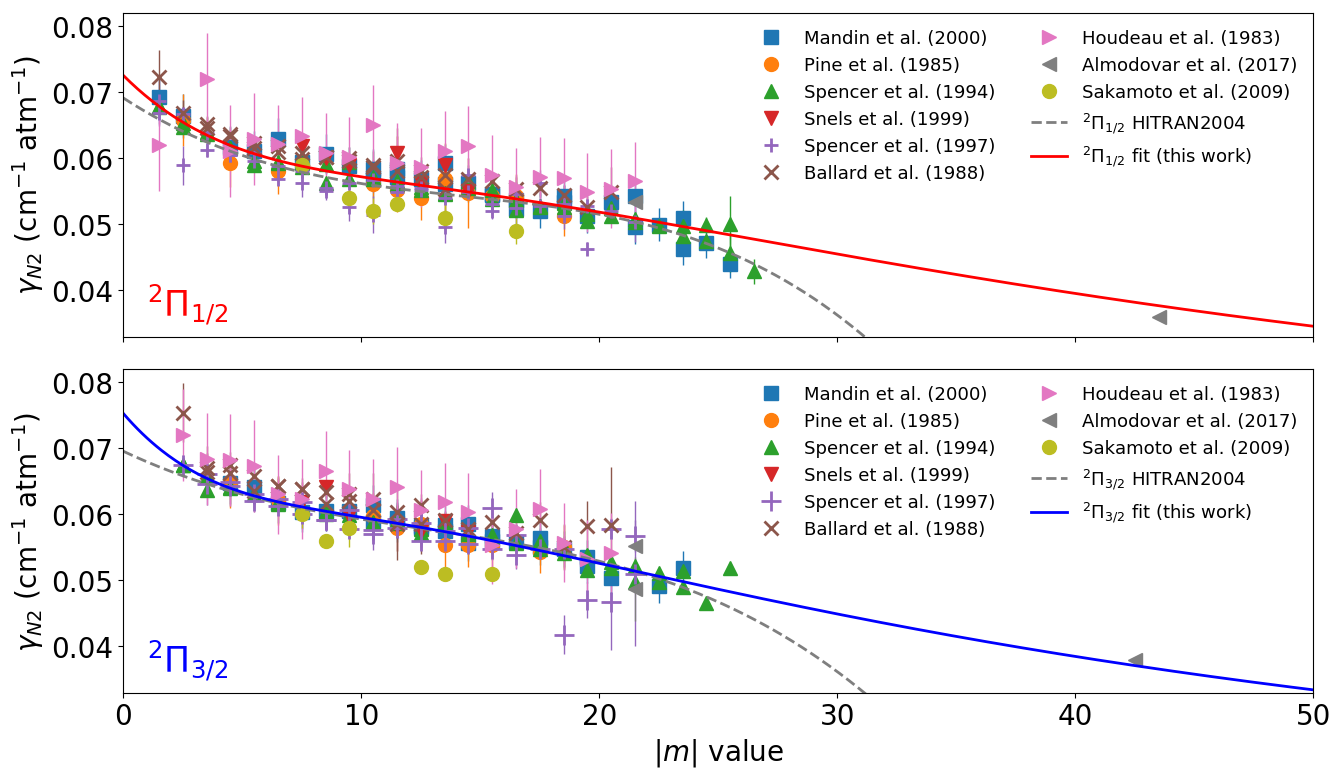}
\caption{N$_{2}$-broadening measurements of NO, for the $1-0$ fundamental band \cite{1994JMoSp.165..506S, 1983JChPh..79.1634H, 1988JMoSp.127...70B, 1997JMoSp.181..307S, 2017JQSRT.203..572A}, and the $2-0$ \cite{1985JMoSp.114..132P, 2000JQSRT..66...93M}, $3-0$ \cite{1999OptCo.159...80S} and $4-0$ \cite{2009ChemLet.38..1000S} overtones. The previous functions used for HITRAN2004 \cite{2005JQSRT..96..139R} have been plotted along with a new fit to the data using Eq.~\ref{eqn_broadening} (this work).\label{no_n2_broad}}
\end{figure}

The air-broadening (Lorentz half width) for NO has been determined by refitting the available data for N$_{2}$- and O$_{2}$-broadening. Fig.~\ref{no_n2_broad} includes the data that were used to obtain the N$_{2}$-broadening parameters. These data encompass the $1-0$ fundamental band \cite{1994JMoSp.165..506S, 1983JChPh..79.1634H, 1988JMoSp.127...70B, 1997JMoSp.181..307S, 2017JQSRT.203..572A}, and the $2-0$ \cite{1985JMoSp.114..132P, 2000JQSRT..66...93M}, $3-0$ \cite{1999OptCo.159...80S} and $4-0$ \cite{2009ChemLet.38..1000S} overtones. The measurements of \citet{2017JQSRT.203..572A} were taken at high temperature (1000~K), which they estimate gives an $\sim$10\% error (when comparing to measurements at 296~K) due to the breakdown of the single power law for large temperature ranges. For this reason, the values from \citet{2017JQSRT.203..572A} have been increased by 10\%. Considering the available data and associated uncertainties, there was no justification for a vibrational dependence to the N$_{2}$-broadening, and all vibrational bands have been included in a least squares weighted fit. A new Pad\'{e} approximation functional form,
\begin{equation}\label{eqn_broadening}
  \gamma (m) = \frac{a_{1} + c_{1}|m|^{2} }{ a_{2} + b_{2}|m| + c_{2}|m|^{2} + d_{2}|m|^{3} } \;,
\end{equation}
\noindent
has been used to provide improved performance for high-$J$ transitions. A separate fit has been used for $\Omega=1/2$ and 3/2 components and the coefficients for N$_{2}$-broadening are given in Table~\ref{tab_no_broad}.

\begin{table*}[t!]
\begin{center}
\caption{Broadening coefficients for NO to be used with Eq.~\ref{eqn_broadening} \label{tab_no_broad} }
\begin{tabular}{lccccccc}
\hline
\vspace{-0.3cm}
 & & \\
Broadener & $\Omega$ &  $a_{1}$             &  $c_{1}$             & $a_{2}$             &  $b_{2}$             & $c_{2}$             & $d_{2}$     \\
\hline
\vspace{-0.3cm}
 & & \\
N$_{2}$   &    1/2   &  6.926$\times10^{-1}$  &  3.312$\times10^{-3}$  & 9.551                 &  4.751$\times10^{-1}$  & 2.244$\times10^{-2}$  &  1.362$\times10^{-3}$  \\\vspace{0.1cm}
N$_{2}$   &    3/2   &  1.159$\times10^{-2}$  &  9.463$\times10^{-5}$  & 1.540$\times10^{-1}$  &  8.590$\times10^{-3}$  & 7.366$\times10^{-4}$  &  3.999$\times10^{-5}$  \\
\hline
\vspace{-0.3cm}
 & & \\
O$_{2}$   &    1/2   &  5.156$\times10^{-1}$  &  2.204$\times10^{-3}$  & 7.739                 &  5.651$\times10^{-1}$  & 1.113$\times10^{-2}$  &  1.174$\times10^{-3}$  \\\vspace{0.1cm}
O$_{2}$   &    3/2   &  1.018$\times10^{-2}$  &  6.262$\times10^{-5}$  & 1.515$\times10^{-1}$  &  9.218$\times10^{-3}$  & 5.476$\times10^{-4}$  &  2.893$\times10^{-5}$  \\
\hline
\vspace{-0.3cm}
 & & \\
Self      &    1/2   &  2.935$\times10^{-1}$  &  1.486$\times10^{-2}$  & 3.561                 &  3.017$\times10^{-1}$  & 2.049$\times10^{-1}$  &  1.745$\times10^{-3}$  \\  \vspace{0.1cm}
Self      &    3/2   &  7.619                 &  8.825$\times10^{-2}$  & 95.45  &  6.114                 & 6.421$\times10^{-1}$  &  3.056$\times10^{-2}$  \\
\hline
\end{tabular}
\end{center}
\end{table*}

\begin{figure}[t!]
\centering
\includegraphics[scale=0.3, trim={0 0 0 0}, clip] {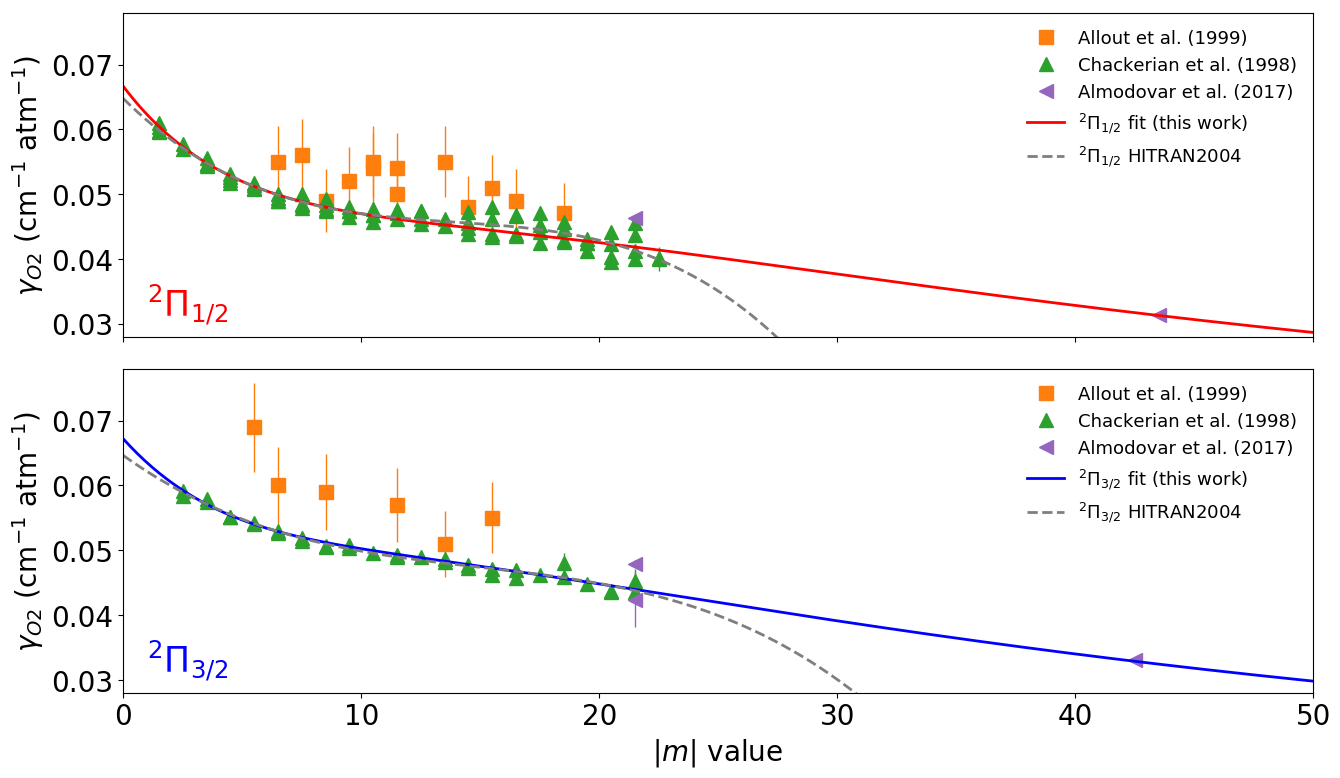}
\caption{O$_{2}$-broadening measurements for NO, for the $1-0$ fundamental band \citep{1998JMoSp.192..215C} and $2-0$ overtone \citep{1999JQSRT..61..759A}. Scaled values from Ref. \citep{1999JQSRT..61..759A} have also been included in the fit. The previous functions used for HITRAN2004 \cite{2005JQSRT..96..139R} have been plotted along with a new fit to the data using Eq.~\ref{eqn_broadening} (this work).\label{no_o2_broad}}
\end{figure}

The O$_{2}$-broadening data include the $1-0$ fundamental band \citep{1998JMoSp.192..215C} and $2-0$ overtone \citep{1999JQSRT..61..759A} (Fig.~\ref{no_o2_broad}). \citet{1998JMoSp.192..215C} have previously observed an  approximate  17\% difference between O$_{2}$- and N$_{2}$-broadening. Therefore, to improve the fitting performance for large $m$, the N$_{2}$-broadening measurements from \citet{2017JQSRT.203..572A} have been scaled by 17\% and included in the fit. The same functional form has been used (Eq.~\ref{eqn_broadening}) and the fitting coefficients for O$_{2}$-broadening are provided in Table~\ref{tab_no_broad}.

Instead of separate N$_{2}$- and O$_{2}$-broadening parameters, HITRAN and HITEMP provide air-broadening for all line-by-line molecules as standard. Hence, the values for O$_{2}$- and N$_{2}$-broadening are combined, such that
\begin{equation}\label{eqn_air_broadening}
  \gamma_{air}(m) = 0.79\gamma_{N_{2}}(m) + 0.21\gamma_{O_{2}}(m)\; .
\end{equation}

\begin{figure}[t!]
\centering
\includegraphics[scale=0.3, trim={0 0 0 0}, clip] {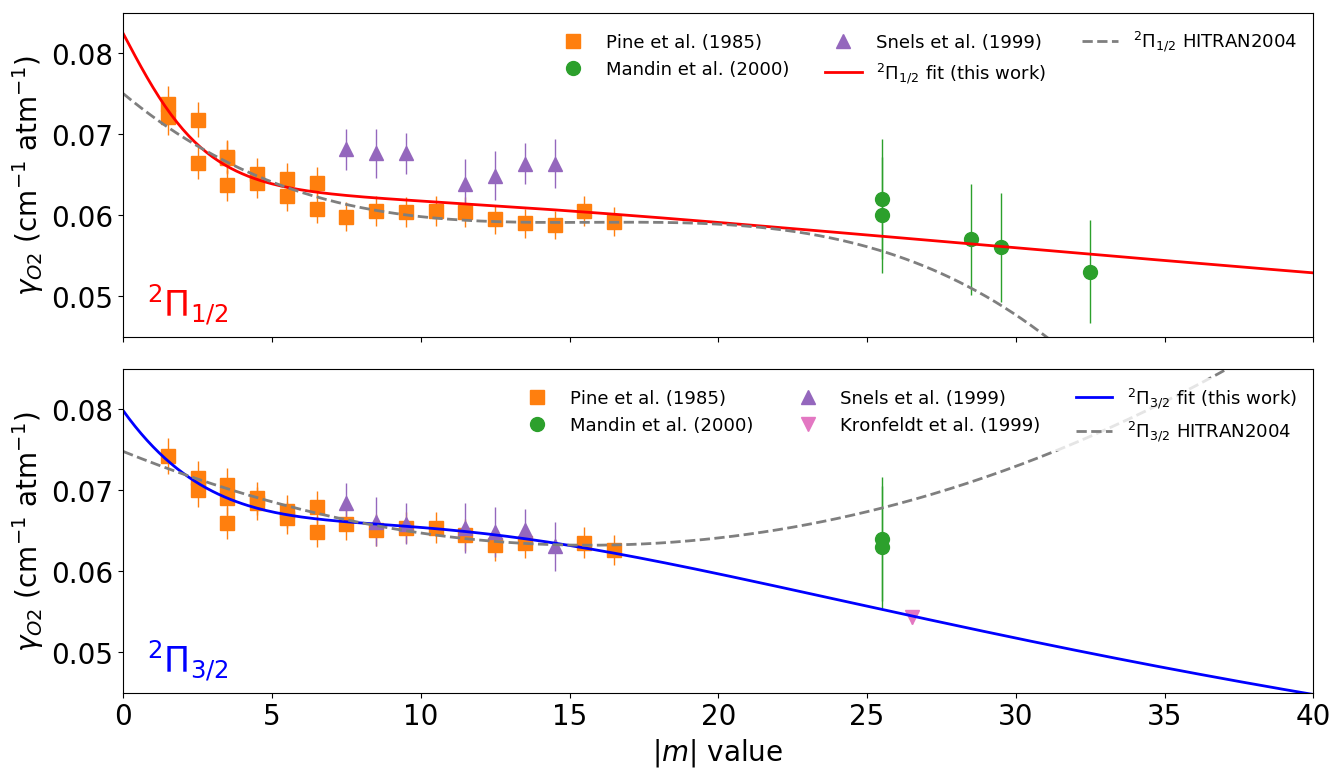}
\caption{Self-broadening measurements for NO, for the $1-0$ fundamental band \citep{1996JOSAB..13.1859K}, and the $2-0$ \cite{1985JMoSp.114..132P, 2000JQSRT..66...93M} and $3-0$ \citep{1999OptCo.159...80S} overtones. The previous functions used for HITRAN2004 \cite{2005JQSRT..96..139R} have been plotted along with a new fit to the data using Eq.~\ref{eqn_broadening} (this work).\label{no_self_broad}}
\end{figure}

Self-broadening for NO has also been revisited. The primary source of self-broadening data comes from the $2-0$ overtone measured by \citet{1985JMoSp.114..132P}, which formed the basis of NO self-broadening in HITRAN2004 \cite{2005JQSRT..96..139R}. These measurements have been combined with additional data for the $2-0$ overtone \citep{2000JQSRT..66...93M} as well as data for the $3-0$ overtone \citep{1999OptCo.159...80S}. A single value for the $^{2}\Pi_{3/2}$ $R(25.5)$ $\Lambda$-doublet of the $1-0$ fundamental has also been included from \citet{1996JOSAB..13.1859K}. The self-broadening data from \citet{1988JMoSp.127...70B} for the  $1-0$ fundamental has been omitted from the fit due to large disagreement with the accepted values of \citet{1985JMoSp.114..132P}. Indeed, \citet{1998JQSRT..60..825G} previously favored the \citet{1985JMoSp.114..132P} measurements over the \citet{1988JMoSp.127...70B} measurements as they provided improved consistency with the results of \citet{1997JMoSp.181..307S}, also used in this work. Again, Eq.~\ref{eqn_broadening} has been used perform a least squares weighted fit of the available data (Fig.~\ref{no_self_broad}), and the fitting coefficients are provided in Table~\ref{tab_no_broad}.

\subsection{Overview of the updated NO line list for HITRAN and HITEMP}\label{no_summary}

The NO line list described in this work has been used to update the HITEMP database. 

Due to the reanalysis of the broadening and pressure shift parameters, the HITRAN database has also been updated. An intensity threshold has been used to limit the number of lines included in the HITRAN update. For any line with $S_{T} > S_{thresh}$ at $T = 100$, 296, 500, 1000 and 2000~K (where $S_{thresh} = 1.0\times 10^{-31}$ cm/molecule), the transition is retained in the HITRAN list. The purpose of this threshold is to keep the NO list complete at the higher temperatures observed in NLTE observations. An additional threshold of $S_{296K} > 1.0\times 10^{-99}$ cm/molecule has been used to avoid a three-digit exponent in the HITRAN ``.par'' format. These thresholds have not been applied to the HITEMP update.

A summary of the updated HITRAN and HITEMP line lists for NO is given in Table~\ref{tab_no_overview}. An overview of the complete HITEMP line list can be seen in Fig.~\ref{no_full_overview}.

\begin{table}[t!]
\begin{center}
\caption{Summary of the updated HITEMP and HITRAN line lists from this work \label{tab_no_overview} }
\begin{tabular}{lcccc}
\hline
\vspace{-0.3cm}
 & & \\
                             &      HITEMP (this work)    &            HITEMP2010 &         HITRAN (this work) &         HITRAN2016   \\
\hline
\vspace{-0.3cm}
 & & \\
Total line count             &               1,137,192    &               115,610 &                 384,305   &               105,079 \\
$^{14}$N$^{16}$O line count  &                 901,528    &               114,232 &                 251,898   &               103,701 \\
$^{15}$N$^{16}$O line count  &                 114,486    &                   699 &                  67,370   &                   699 \\
$^{14}$N$^{18}$O line count  &                 118,178    &                   679 &                  65,037   &                   679 \\
$J_{max}''$                  &                   184.5    &                 125.5 &                   156.5   &                 125.5 \\
$v_{max}'$                   &                      29    &                    14 &                      26   &                    14 \\
$\Delta v_{max}$             &                      16    &                     5 &                      14   &                     5 \\
$\nu_{min}$ (cm$^{-1}$)      &     1.00$\times10^{-6}$    &   1.00$\times10^{-6}$ &      1.00$\times10^{-6}$  &   1.00$\times10^{-6}$ \\
$\nu_{max}$ (cm$^{-1}$)      &                  26,777    &                 9,273 &                   23,727  &                 9,273 \\
$S_{min}$ (cm/molecule)      &   1.72$\times10^{-138}$    & 2.18$\times10^{-121}$ &     1.00$\times10^{-99}$  &  1.45$\times10^{-95}$ \\  \vspace{0.1cm}
$S_{max}$ (cm/molecule)      &    2.32$\times10^{-20}$    &  2.32$\times10^{-20}$ &     2.32$\times10^{-20}$  &  2.32$\times10^{-20}$ \\
\hline
\end{tabular}
\end{center}
\end{table}



\section{Nitrogen Dioxide (molecule 10)}\label{mol_no2}

\subsection{Description of NO$_{2}$ line lists}\label{line_lists_no2}
NO$_{2}$ was not included as part of HITEMP2010 \cite{2010JQSRT.111.2139R}. A new line list of NO$_{2}$ has been completed for this work using current line parameters contained in HITRAN2016 \citep{2017JQSRT.203....3G} and the recent \textit{NDSD-1000} \citep{2016JQSRT.184..205L} line list. These line lists are briefly described below.

\subsubsection{NO$_{2}$ in HITRAN2016}\label{list_hitran_no2}
The NO$_{2}$ line parameters in HITRAN2016 include data for the principal isotopologue ($^{14}$N$^{16}$O$_{2}$) with a natural abundance of 0.991616. This constitutes 104,223 lines in the region 0 -- 3075.0~cm$^{-1}$ with intensities $S > 1.0\times10^{-25}$ cm/molecule (see \citet{1998JQSRT..60..839P} for a detailed description). The data encompass three infrared regions dominated by the $\nu_{2}$,  $\nu_{3}$ and  $\nu_{1}+\nu_{3}$ bands  (13.3, 6.2 and 3.4~$\mu$m, respectively), in addition to pure rotational transitions. All regions contain spin-rotation interactions ($J = N \pm 1/2$), but pure rotation and $\nu_{2}$ fundamental transitions also include additional hyperfine structure.

\subsubsection{The \textit{NDSD-1000}}\label{list_ndsd_no2}

The Nitrous Dioxide Spectroscopic Databank at 1000~K (\textit{NDSD-1000})  contains data of the principal isotopologue of NO$_{2}$ in the region 466 -- 4476~cm$^{-1}$, and is intended for simulating spectra of NO$_{2}$ up to a temperature of 1000~K  \cite{2016JQSRT.184..205L}. The databank has been constructed from a global model of line intensities and positions, through the application of an effective Hamiltonian and dipole operator approach, with the inclusion of spin-rotation interaction ($J = N \pm 1/2$). This method produces line positions with a root mean square residual of 0.015~cm$^{-1}$ when compared to observations, which can be as low as 0.001~cm$^{-1}$ for fundamental bands. In addition, the databank also provides broadening coefficients and temperature-dependent exponents at 296 and 1000~K. \textit{NDSD-1000} is available as five separate files\footnote{\href{ftp://ftp.iao.ru/pub/NDSD/}{ftp://ftp.iao.ru/pub/NDSD/}} separated by polyad bands for $\Delta P=1$, 2, 3, 4 and 6 (where $P = 2V_{1}+V_{2}+2V_{3}$, and $V_{1}$, $V_{2}$, and $V_{3}$ are the quantum numbers of the normal modes of vibration), with a total of  1,051,689 lines.

\begin{figure}[t!]
\centering
\includegraphics[scale=0.3, trim={0 0 0 0}, clip] {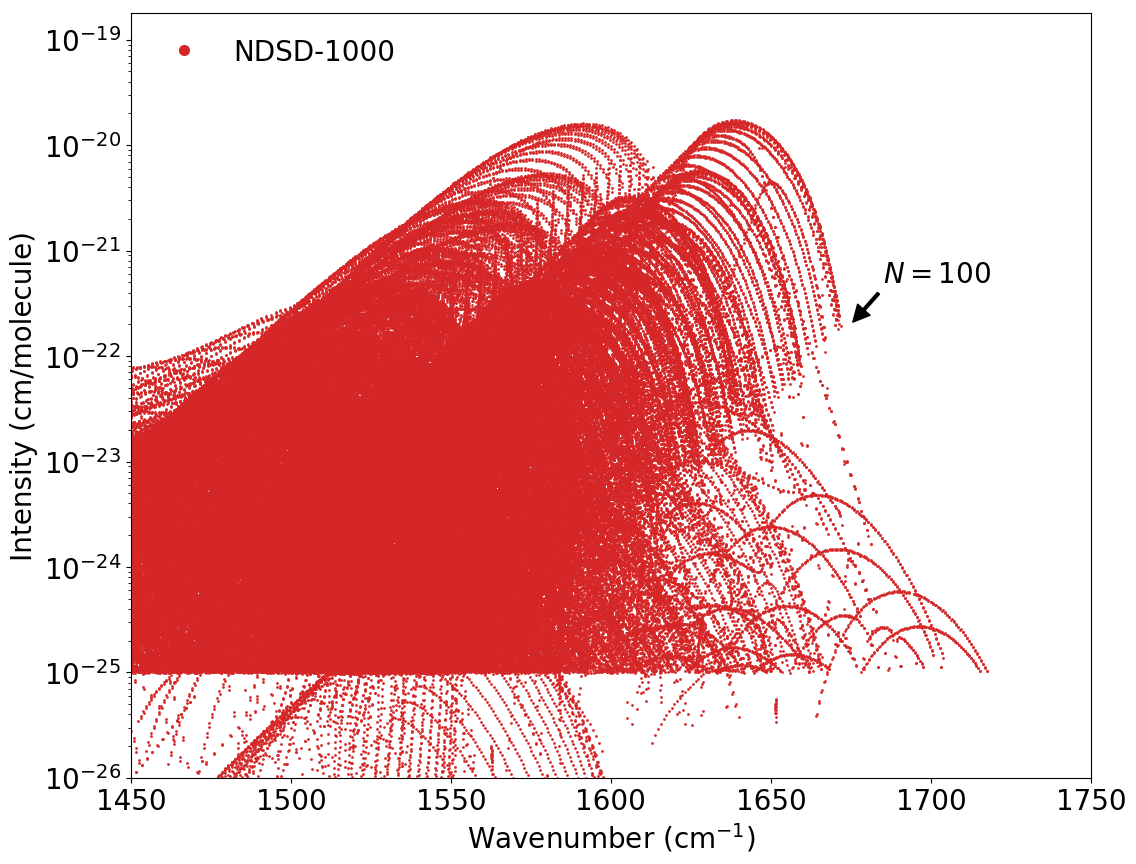}
\caption{A portion of the \textit{NDSD-1000} line list with intensities shown at 1000~K. The intensity cutoff at $S = 1.0\times10^{-25}$ cm/molecule is clearly seen. Lines with an intensity (at 1000~K) below this cutoff, exceed this intensity at 296~K. Also shown is a limitation caused by the high rotational level restriction, as indicated by the missing lines beyond $N = 100$ for the $\nu_{3}$ bands. \label{ndsd_high_rot}}
\end{figure}

During the construction of the databank, only transitions with an intensity $S > 1.0\times10^{-25}$ cm/molecule at 296 or 1000~K have been included. In addition, restrictions were also applied to the maximum rotation levels ($N \leq 100$ and  $K_{a}\leq 20$). This is because of a poor convergence of the polynomial expansion for both effective operators (Hamiltonian and electric dipole) due to the limited coverage of the input data ($N \leq 62$ and  $K_{a} \leq 12$) \cite{2016JQSRT.184..205L}. This was to retain accuracy for high resolution applications; however these excluded lines would be suitable for low resolution comparisons. Where possible, lower-state energy levels have been taken from HITRAN2016 for accuracy. It should be noted that the restrictions applied to the maximum rotation levels will limit the completeness at higher temperature regimes. For example, Fig.~\ref{ndsd_high_rot} shows how the population of these higher rotation levels becomes relatively strong at 1000~K for $\nu_{3}$ region.

\begin{figure}[t!]
\centering
\includegraphics[scale=0.3, trim={0 0 0 0}, clip] {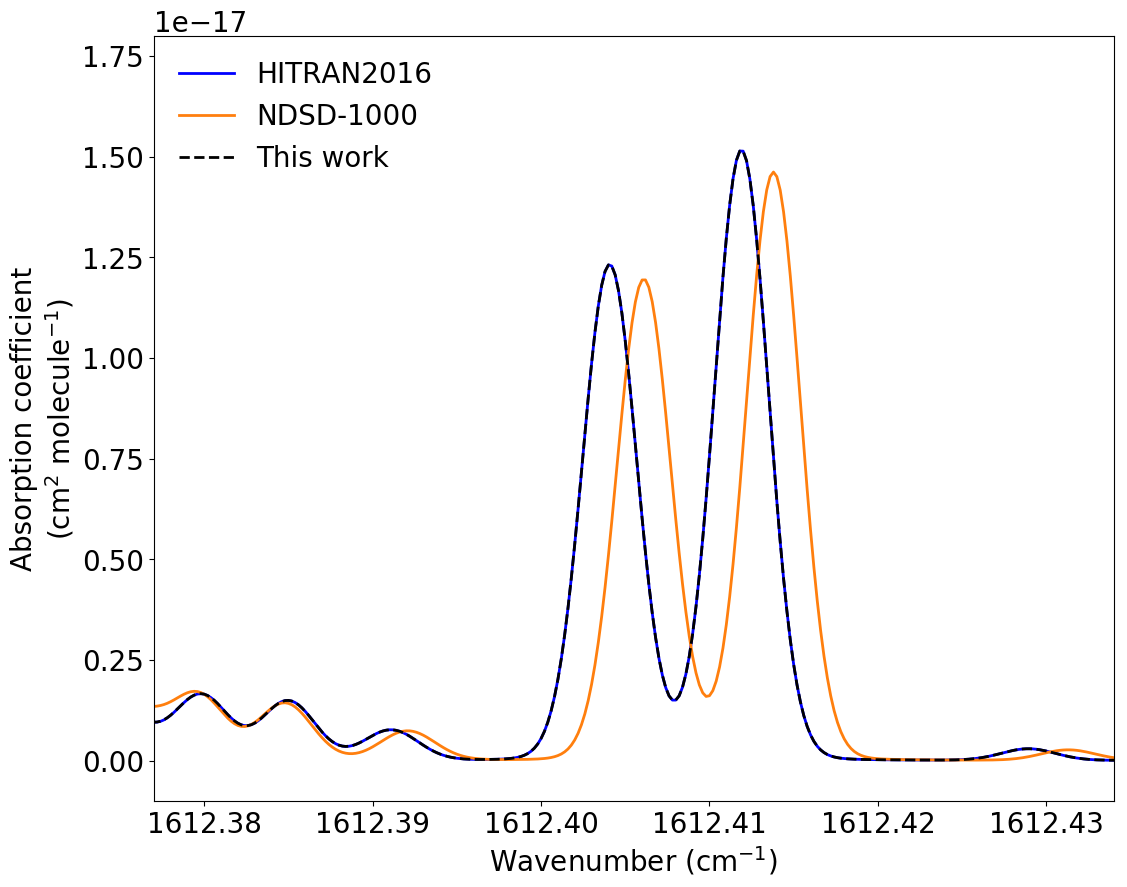} \\
\includegraphics[scale=0.3, trim={0 0 0 0}, clip] {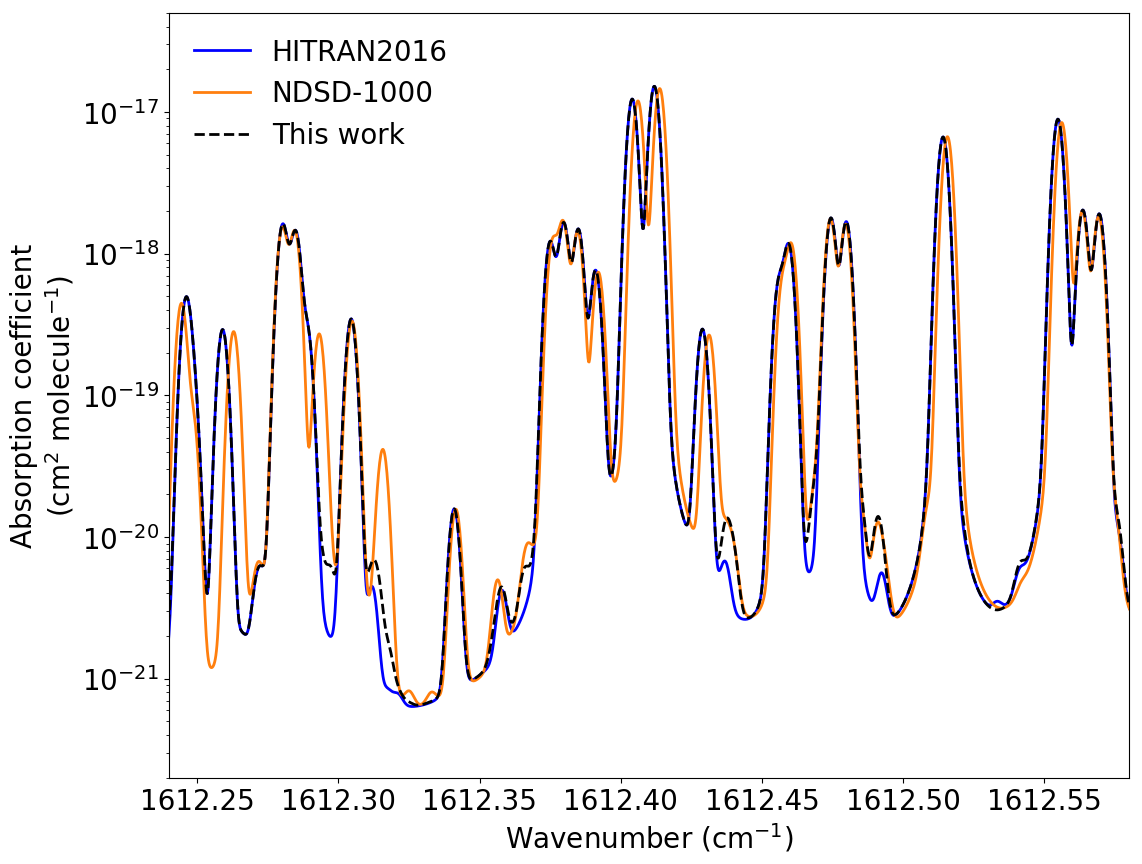}
\caption{Calculated absorption cross sections using HITRAN2016 and NOSD-1000 line lists, and this work for a portion of the $\nu_{3}$ band of NO$_{2}$ near 1612.4 cm$^{-1}$. The upper panel is comparable to Fig.~2 in Ref. \cite{2017JQSRT.200...12P} and shows the improved intensities of the recalculated NOSD-1000 line list \cite{2017JQSRT.202...37L}. However a slight wavenumber shift remains (when compared to HITRAN2016). The lower panel shows the same calculated cross sections but with a logarithmic scale. This is used to highlight differences between each line list and this work. All spectra have been calculated using HAPI \cite{2016JQSRT.177...15K}. \label{ndsd_int_pos}}
\end{figure}

A recent article by \citet{2017JQSRT.200...12P} highlighted issues with the intensities of the original \textit{NDSD-1000} line list. Intensities between the $J = N + 1/2$ and $J = N - 1/2$ electron spin rotation components were shown to be inconsistent with high-resolution experimental measurements and previous literature values (including HITRAN2016). Consequently, the line intensities were recalculated \cite{2017JQSRT.202...37L}, and Fig.~\ref{ndsd_int_pos} shows the improvement for intensity when compared to HITRAN2016. The recalculated intensities from  \citet{2017JQSRT.202...37L} are used for the update described in this work.

\begin{figure}[t!]
\centering
\includegraphics[scale=0.3, trim={0 0 0 0}, clip] {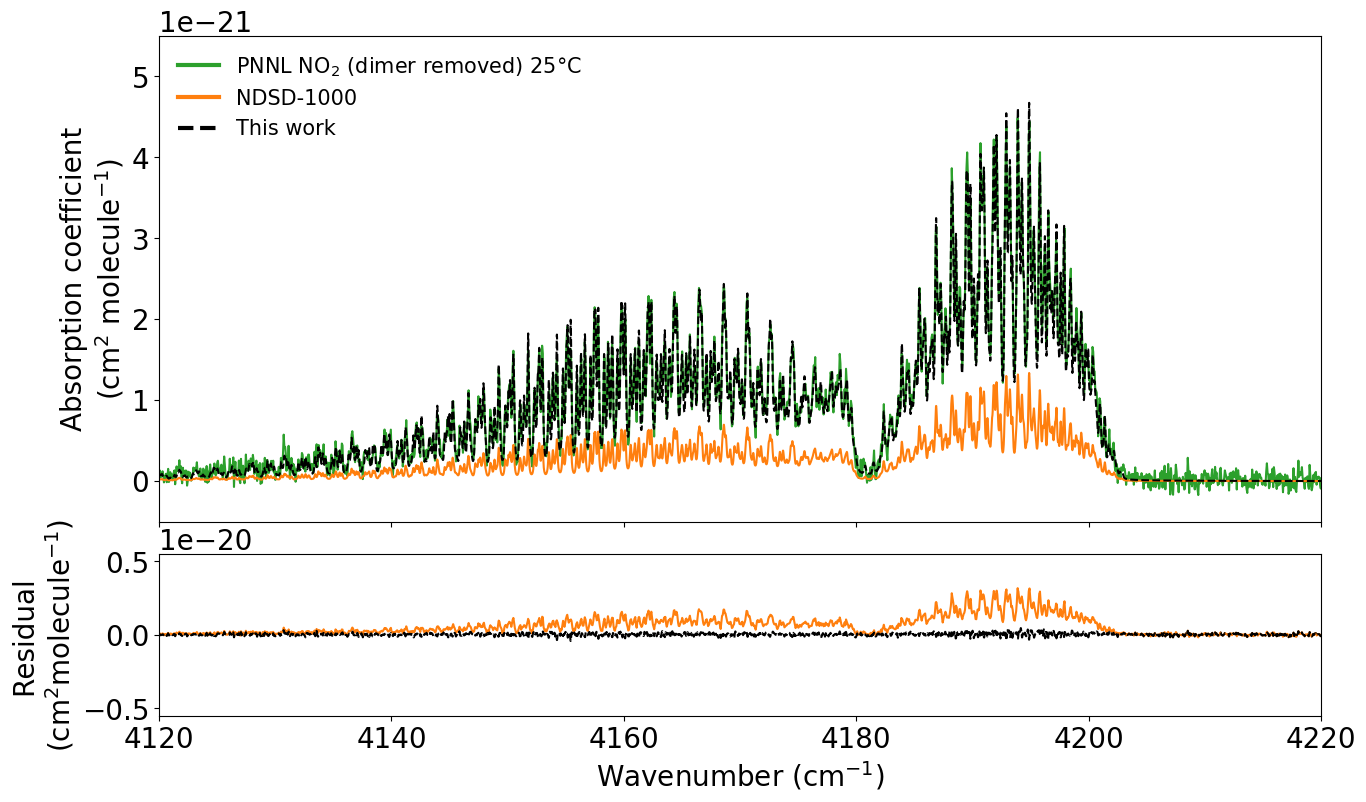}
\caption{PNNL absorption cross sections at 25$^{\circ}$C compared to calcualted cross sections using the \textit{NDSD-1000} line list and this work. A residual of PNNL--calculation is shown in the lower panel. This work contains $\Delta P=6$ lines from \textit{NDSD-1000}  multiplied by a scale factor of 3.5. All spectra have been calculated using HAPI \cite{2016JQSRT.177...15K}.\label{ndsd_p6}}
\end{figure}

\citet{2016JQSRT.184..205L} compared PNNL absorption cross sections \cite{2004ApSpe..58.1452S} of NO$_{2}$ to those based on the \textit{NDSD-1000} line list for the spectral regions containing $\Delta P=1$, 2 and 4 polyad bands. Fig.~\ref{ndsd_p6} shows a similar comparison for the $2\nu_{1}+\nu_{3}$ band (part of $\Delta P=6$) where there are no data included in HITRAN2016. It was apparent that the \textit{NDSD-1000} line list was approximately 3.5 times weaker than the PNNL observations for the  $\Delta P=6$ region. Hence, all $\Delta P=6$ lines from  \textit{NDSD-1000} have  been scaled by a factor or 3.5 for this work.

It should be noted that integrated line intensities for the $3\nu_{3}$ band ($\Delta P=6$) provided by \citet{2000JMoSp.201..134S} is in excellent agreement with \textit{NDSD-1000} (before the scale factor is applied). This is because the \textit{NDSD-1000} line intensities are based on the measurements of \citet{2000JMoSp.201..134S}.  Excellent agreement is also seen between integrated line intensities of bands in the $\Delta P=4$ region. However, only a scale factor was necessary for $\Delta P=6$ bands. The relative band intensities of the PNNL cross sections are reliable because the $500 - 7000$ cm$^{-1}$ spectral region is recorded as a single measurement. This led us to the conclusion to give preference to the PNNL data. 

\subsection{Combining line lists of NO$_{2}$}\label{list_ndsd_combined}

\citet{2017JQSRT.200...12P} demonstrated that the line intensities and positions of HITRAN2016 were excellent at reproducing high-resolution Fourier transform spectral measurements of $\nu_{3}$ line intensities. Similar accuracy has also previously been demonstrated for the other data included in HITRAN2016 \cite{1988MolPh..63..791P, 1993JMoSp.160..456P, 1997JMoSp.181..379M, 1998JQSRT..60..839P}. While the line intensities of the \textit{NDSD-1000} line list have been corrected, the positions and intensities of HITRAN2016 are generally superior. HITRAN2016 also includes hyperfine structure. For these reasons, the \textit{NDSD-1000} line list has been used to supplement the HITRAN2016 data when producing the NO$_{2}$ line list for this work.

The five polyad regions of the \textit{NDSD-1000} line lists have been combined and converted into the standard HITRAN format \cite{2017JQSRT.203....3G}. The line parameters for HITRAN and HITEMP are provided at a reference temperature of 296~K. Therefore the intensities, air- and self-broadening, and temperature exponents calculated at 296~K have been retained from the \textit{NDSD-1000} line list.

All transitions from \textit{NDSD-1000} that are not included in HITRAN have been combined with the HITRAN2016 line list to create an NO$_{2}$ line list for this work. This line list includes numerous hot and combination bands that were not part of HITRAN2016, with an increased spectral coverage from 0 -- 4476~cm$^{-1}$.

Intensity comparisons between lines lists show good agreement. However for some small regions with large $N$ and $K_{a}$  (e.g., for the $\nu_{1}$ band), the intensities of  \textit{NDSD-1000} begin to diverge from HITRAN2016. This is a consequence of the differing methods for the treatment of intensities. In the case of HITRAN, the effective Hamiltonians have been treated for a single or group of interacting vibrational states, as opposed to the global approach used for \textit{NDSD-1000}. This will lead to slight intensity discontinuities at the merging points in the line list of this work. However, since $S_{min} = 1.0 \times 10^{-25}$ cm/molecule for HITRAN2016, these discontinuities occur at weak regions of the NO$_{2}$ spectrum.

Recently, \citet{Vasilenko2018} reported on comparisons between \textit{NDSD-1000} and HITRAN2016 line lists and their potential influence on remote sensing applications. They have observed that the spectral parameters in two databases are different, however, it was erroneously concluded that HITRAN2016 was incorrect. In reality for the lines that are sufficiently strong in the range of temperatures encountered in the terrestrial atmospheres, the line intensities and positions in HITRAN2016 are superior to those from  \textit{NDSD-1000} (as outlined above). With that being said some weak lines and bands are indeed missing in HITRAN2016 and it makes sense to include them here. 

\begin{figure}[t!]
\centering
\includegraphics[scale=0.3, trim={0 0 0 0}, clip] {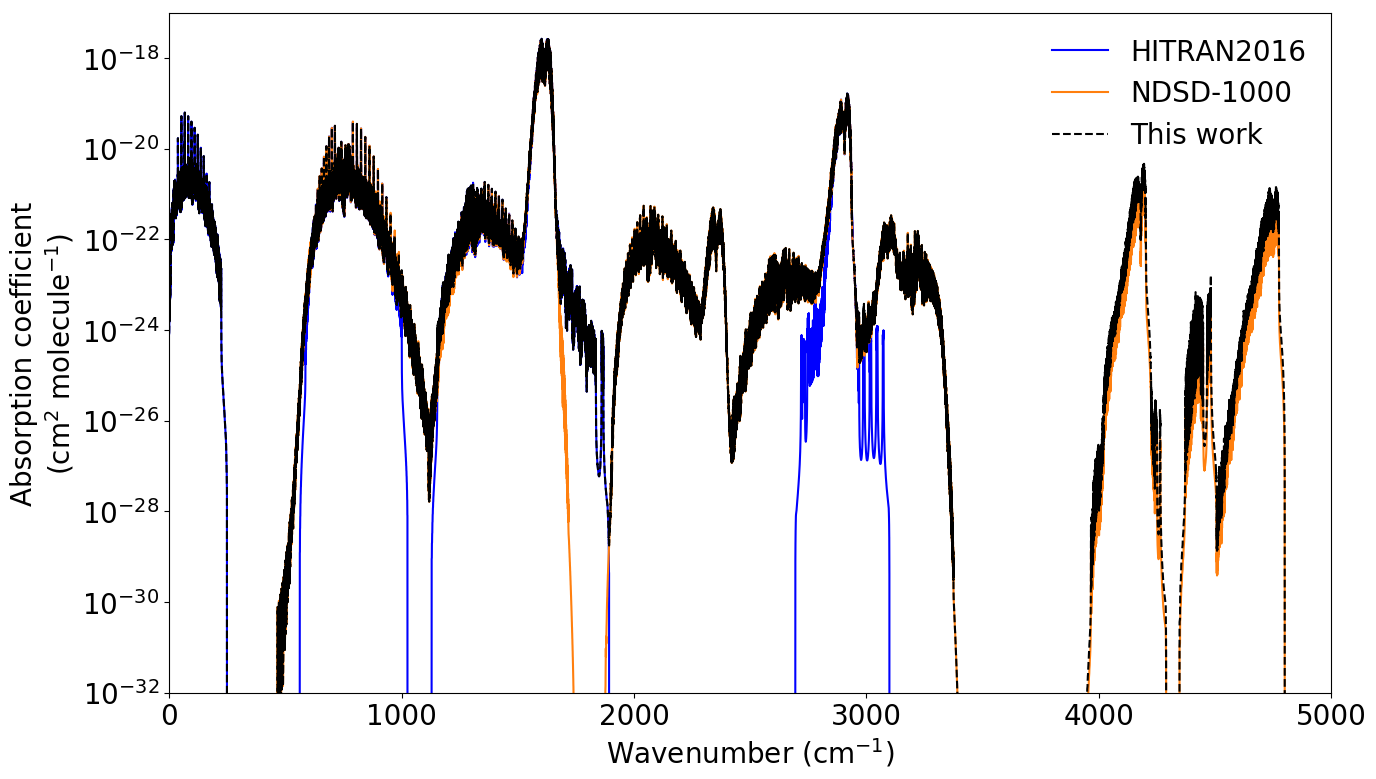}
\caption{Calculated absorption cross sections at 296~K using line lists from HITRAN2016, \textit{NDSD-1000} and this work. All spectra have been calculated using HAPI \cite{2016JQSRT.177...15K}.\label{ndsd_xsc}}
\end{figure}

One benefit of including all HITRAN2016 NO$_{2}$ lines into this work is that comparisons will agree with previous work. Fig.~\ref{ndsd_int_pos} contains a small spectral region highlighted by \citet{2017JQSRT.200...12P} to demonstrate the differing intensities of HITRAN2016 and \textit{NDSD-1000}. Here the \textit{NDSD-1000} intensities are shown to have been improved; however the wavenumber shifts remain. It can be seen that this work matches HITRAN2016 almost exactly. A logarithmic scale has also been used to indicate where this work and HITRAN2016 differ. Furthermore, Fig.~\ref{ndsd_p6} also includes a cross section calculated using this work, where a clear improvement over \textit{NDSD-1000} is seen when compared to PNNL. The full spectral range is shown in Fig.~\ref{ndsd_xsc}, and this work is seen to provide the most compete data set for the spectral region up to 4476~cm$^{-1}$.

\subsubsection{Recalculation of pressure shifts for NO$_{2}$ and updates to HITRAN}\label{list_ndsd_shifts}

Pressure-induced line shifts for air ($\delta_{air}$) have also been included for this work. The pressure shifts from HITRAN2016 are based on multi-spectrum measurements of the $\nu_{3}$ fundamental band by \citet{2004JMoSp.228..593B}. These shifts depend on $m$ (where $m=N''$ for $P$- and $Q$-branches and $m=N''+1$ for $P$-branches) and have been fit to $K_{a}''$ dependent quadratic functions (up to $K_{a}''=7$).

To expand these $\nu_{3}$ pressure-shift functions ($\delta_{001}(m,K_{a})$, where 001 refers to $\nu_{3}$) to other vibrational bands, a scaling factor has been applied as was done for HITRAN2004 \cite{2005JQSRT..96..139R}. This factor is based on the ratio of approximate band centers such that the new pressure-induced shift,
\begin{equation}\label{eqn_shifts_ndsd}
 \delta(m,K_{a}) = \frac{ \Delta n_{1}\omega_{1} + \Delta n_{2}\omega_{2} + \Delta n_{3}\omega_{3} }{ \omega_{3} }  \delta_{001}(m,K_{a})  \; ,
\end{equation}
where $\omega$ refers to the band center of each fundamental band ($\omega_{1} = 1319.766$,  $\omega_{2} = 749.653$, and $\omega_{3} = 1616.849$~cm$^{-1}$) and $\Delta n$ is the change in quanta for each vibrational band between upper and lower states. For transitions with $K_{a}''>7$, the values of $K_{a}''=7$ have been used, and $m$ has been limited to the maximum measured values of \citet{2004JMoSp.228..593B} for each $K_{a}''$.

While calculating the pressure shifts for the this work, it became apparent that the pressure shifts in HITRAN2016 for NO$_{2}$ contained a number of errors. For consistency, Eq.~\ref{eqn_shifts_ndsd} has been used to recalculate the pressure shifts of lines that were included from HITRAN2016. Furthermore, HITRAN2016 contains no $\Delta P=3$ or 6 lines. These bands are observed in comparisons with PNNL spectra (e.g.,  Fig.~\ref{ndsd_p6}), and therefore the $\Delta P=3$ and $\Delta P=6$ lines (with scaled intensities described above) have been used to update HITRAN for lines with $S > 1.0 \times 10^{-25}$ cm/molecule. The updated NO$_{2}$ HITRAN line list now contains 125,193 lines, an increase of 20,970 lines when compared to HITRAN2016.

\subsection{Overview of the NO$_{2}$ line list for HITEMP}\label{no2_summary}

A summary of the NO$_{2}$ line list produced for this work and used to update HITEMP  is given in Table~\ref{tab_no2_overview}, alongside a summary of HITRAN2016 and the additional HITRAN update based on this work. A full cross section overview of the this work  is shown in Fig.~\ref{ndsd_xsc}.

\begin{table*}[t!]
\begin{center}
\caption{Summary of the HITEMP and HITRAN updates of NO$_{2}$ from this work \label{tab_no2_overview} }
\begin{tabular}{lccc}
\hline
\vspace{-0.3cm}
 & & \\
                             &              HITEMP (this work)     &              HITRAN (this work) &              HITRAN2016 \\
\hline
\vspace{-0.3cm}
 & & \\
Total line count             &               1,108,709     &                 125,193 &                 104,223 \\
Vibrational band count       &                     261     &                      31 &                      11 \\
$\nu_{min}$ (cm$^{-1}$)      &                   0.498     &                   0.498 &                   0.498 \\
$\nu_{max}$ (cm$^{-1}$)      &                4775.320     &                4775.320 &                3074.153 \\
$S_{min}$ (cm/molecule)      &    2.29$\times10^{-47}$     &    4.24$\times10^{-28}$ &    4.24$\times10^{-28}$ \\  \vspace{0.1cm}
$S_{max}$ (cm/molecule)      &    1.30$\times10^{-19}$     &    1.30$\times10^{-19}$ &    1.30$\times10^{-19}$ \\
\hline
\end{tabular}
\end{center}
\end{table*}


\section{Conclusions}

High-temperature line lists of NO$_{x}$ (i.e., NO and NO$_{2}$) and N$_{2}$O have been compiled, which (along with the CO line list from \citep{2015ApJS..216...15L}) marks the first updates to the HITEMP database since \citet{2010JQSRT.111.2139R}. In addition to these high-temperature line lists, updates to the HITRAN database for NO and NO$_{2}$ have also been performed. In the case of NO the spectral range in HITRAN is being extended, while intensities of some of the bands have been improved, as well as the intensity ratios between isotopologues. In the case of NO$_{2}$ new bands have been added extending into NIR. The HITRAN updates are already available through the HITRAN website (\href{https://hitran.org}{https://hitran.org}). This work also demonstrates the necessary validation steps that are required before data are included as part of the HITRAN and HITEMP databases. We anticipate numerous uses for the line lists presented here, including combustion monitoring, terrestrial atmospheric observations, breath analyses, and astrophysics applications.

It is well understood that using HITRAN data for applications outside of the appropriate temperature and wavenumber ranges can lead to issues of incompleteness. HITEMP data are no different. The appropriate temperature ranges for each molecule in HITEMP depends on the underlying data used to populate each line list. The HITRAN and HITEMP data can both be used at 296~K; however we stress that for elevated temperatures, only HITEMP should be used. For N$_{2}$O and NO$_{2}$, the majority of lines have been provided by \textit{NOSD-1000} and \textit{NDSD-1000} line lists, respectively, which suggest a maximum applicable temperature of 1000~K. In practice, the maximum applicable temperature (for completeness) may be below this value, particularly for NO$_{2}$, due to restrictions on maximum rotational levels (and missing polyad bands). We therefore advise caution when using the HITEMP N$_{2}$O and NO$_{2}$ data at temperatures approaching this temperature limit. For NO,  \citet{2017MNRAS.470..882W} suggest a maximum applicable temperature of 5000~K (for the \textit{NOname} line list). In forming the HITEMP line list for NO for this work, data have been excluded because of doubts regarding accuracy for high rotational levels and vibrational bands. It is therefore expected that the maximum temperature range is also reduced. While it is difficult to place a value for the maximum applicable temperature for this work (due to merging and data restrictions), we advise caution when using this line list for temperatures beyond $\sim$4000~K. If the complete versions of each line lists are required, we refer the reader to the reference sources.

In addition to line intensities, the broadening, temperature exponents and pressure shift parameters also depend on temperature. For HITEMP (and HITRAN), these parameters are provided at the reference temperature of 296~K. However, the uncertainties for these parameters will increase as the temperature increases. Adding numerous parameters at specific temperature ranges is not practical for the HITEMP database, but new formalisms have been proposed \cite{Gamache2018440} that will be considered for future editions of HITEMP. If additional parameters are required (that were not included in this work), we again refer the reader to the reference sources.

The PNNL infrared absorption cross-section archive is an excellent source for room temperature comparisons. However the relative low spectral resolution ($\sim$0.1~cm$^{-1}$) is insufficient for high-resolution comparisons. One difficulty when validating line lists is a lack of high-resolution spectra available in the literature, particularly at high temperatures. The point has recently been emphasized by \citet{2016arXiv160206305F} when considering the data requirements necessary for interpreting future exoplanet atmospheric observations.

This work expands the HITEMP database to include seven molecules. N$_{2}$O and NO$_{2}$ are new additions and they join H$_{2}$O, CO$_{2}$, CO and OH, with NO being updated. Recent advancements in theoretical and semi-empirical methods have increased the reliability of calculated line lists. Further improvements are therefore planned for HITEMP over the coming month/years, these include the addition of new molecules (CH$_{4}$, SO$_{2}$, NH$_{3}$, HCN, C$_{2}$H$_{2}$, PH$_{3}$, C$_{2}$, CH, CN and H$_{3}^{+}$), additions adapted from HITRAN2016 (HF, HCl, HBr, HI and H$_{2}$), as well as updates for H$_{2}$O, CO$_{2}$, CO and OH.

Access to both the HITRAN and HITEMP databases is freely available through the HITRAN website (\href{https://hitran.org}{https://hitran.org}). The considerable size of the HITEMP line lists mean that these data are available to download as static files. In the near future a more flexible structure will be introduced. The HAPI libraries are undergoing significant updates, which enable a substantial speed improvement for calculations; thereby allowing HAPI to carry out line-by-line calculations on large HITEMP linelists. The HITRAN data are available via a more flexible SQL structure, and can be downloaded from the website or through the HAPI software (see Ref. \cite{2016JQSRT.177...15K} for more details).

\section{Acknowledgements}
We would like to acknowledge N. N.
Lavrentieva and A. C. Dudaryonok for their work in the production of the \textit{NOSD-1000} and \textit{NDSD-1000} line lists, as well as A. Wong for work on the \textit{NOname} line list.
Update of the HITRAN and HITEMP databases was supported through the NASA Aura and PDART grants NNX17AI78G and NNX16AG51G, respectively.

\section{References}

\bibliography{manuscript_jan19_bibfile}

\clearpage

\appendix

\section{Uncertainty codes for NO line lists}\label{app_err_no}
\subsection{CDMS}
NO uncertainty codes (UCs) have been provided in the CDMS line lists. A maximum uncertainty code $1.0\times10^{-8} \leq \nu \leq 1.0\times10^{-7}$ cm$^{-1}$ (i.e., following the HITRAN convention, UC=8) is used for line positions. The applied intensity uncertainty is dependent on $J$ such that for $J \leq 60$ the uncertainty for intensity is 5-10\% (UC=5), and for $J > 60$ the uncertainty for intensity is 10-20\% (UC=4).

\subsection{\textit{NOname}}
Wavenumber uncertainty codes for the \textit{NOname} line list have been deduced from the vibration and rotational level assignment. For $J'' \leq 60.5$ and $10 < v'\leq 20$ an uncertainty of $0.01 \leq \nu \leq 0.1$ cm$^{-1}$ (UC=2) is applied, for $5 < v'\leq 10$ an uncertainty of  $1.0\times10^{-3} \leq \nu \leq 0.01$ cm$^{-1}$ (UC=3), and for $v'\leq 5$ an uncertainty of $1.0\times10^{-4} \leq \nu \leq 1.0\times10^{-3}$ cm$^{-1}$ (UC=4). For $60.5 < J'' \leq 99.5$ and $20 < v' \leq 29$ an uncertainty of $0.1 \leq \nu \leq 1.0$ cm$^{-1}$ (UC=1) is applied. For all other cases, an UC=0 is given for the wavenumber. Intensity uncertainty codes are also based on the vibration and rotational levels. For $J'' \leq 99.5$, an uncertainty of 10-20\% (UC=4) is given for $5 < \Delta v \leq 29$; 5-10\% (UC=5) for $1 < \Delta v \leq 5$; and 2-5\% (UC=6) for $\Delta v \leq 1$. In all other cases an intensity UC=2 is given.

\subsection{HSPM2018}
An uncertainty of $\geq 1.0$ cm$^{-1}$ (UC=0) has been provided for positions, with an uncertainty code applied for estimated intensities (UC=2).

\section{Pressure shifts of NO}\label{app_pressure_no}
The pressures shifts applied to the NO line list have been limited to $|m|=21.5$ due to the increase in uncertainty of the measurements beyond this value. For pressures shifts with $m>21.5$, the value at $m=21.5$ is used. Similarly, for $m<-21.5$, the value at $m=-21.5$ is used. The measurements also do not include some of the smallest $m$ values; in these cases the lowest measured value is applied. It should also be noted that the pressure shifts for the $\Delta v=0$ have been fixed to 0 cm$^{-1}$atm$^{-1}$. While it is possible to extrapolate Eq.~\ref{eqn_shifts} for the $\Delta v=0$ bands, a number of transitions would actually be shifted into the negative wavenumber regime for ambient pressures, and therefore to avoid this issue a value of zero is given.  

\end{document}